\documentclass[
	a4paper, 
	10pt, 
	unnumberedsections, 
	twoside, 
]{LTJournalArticle}

\makeatletter
\newcommand\notsotiny{\@setfontsize\notsotiny\@vipt\@viipt}
\makeatother

\runninghead{XPS Study of Medium-Temperature Baking of Niobium. A.Prudnikava et al.} 

\footertext{\textit{Preprint ver.2, Feb 2024 }} 

\usepackage{siunitx}
\DeclareSIUnit\bar{bar}

\DeclareSIUnit{\atpercent}{at\%}

\usepackage[dvipsnames]{xcolor}

\usepackage[switch]{lineno} 

\usepackage{multirow}

\title{\textit{In-situ} Synchrotron X-Ray Photoelectron Spectroscopy Study of Medium-Temperature Baking of Niobium for SRF Application}

\author{%
	A. Prudnikava\textsuperscript{1}\thanks{Corresponding author: \href{mailto:alena.prudnikava@helmholtz-berlin.de}{alena.prudnikava@helmholtz-berlin.de}},
 Y. Tamashevich\textsuperscript{1,}\thanks{Presently with the Department of Inorganic Spectroscopy, The Max Planck Institute for Chemical Energy Conversion, Stiftstr. 34-36, 45470 Mülheim an der Ruhr, Germany},
 A. Makarova\textsuperscript{2},
 D. Smirnov\textsuperscript{3},\\
 J. Knobloch\textsuperscript{1,4}\\ 
}

\date{\footnotesize\textsuperscript{
\textbf{1}}Helmholtz Centre for Materials and Energy, Albert-Einstein-Str. 15, 12489 Berlin, Germany\\
\textsuperscript{\textbf{2}}Institute of Chemistry and Biochemistry, Freie Universität Berlin, Arnimallee 22, 14195 Berlin, Germany\\
\textsuperscript{\textbf{3}}Institute of Solid State and Materials Physics, Technische Universität Dresden, Haeckelstraße 3, 01069 Dresden, Germany \\
\textsuperscript{\textbf{4}}Department of Physics, Universität Siegen, Walter-Flex-Str. 3, 57068 Siegen, Germany
}

\renewcommand{\maketitlehookd}{%
	\begin{abstract}
		\noindent 
        In order to determine optimal parameters of vacuum thermal processing of superconducting radiofrequency niobium cavities exhaustive information on the initial chemical state of niobium and its modification upon a vacuum heat treatment is required.
In the present work the chemical composition of the niobium surface upon ultra-high vacuum baking at \SIrange{200}{400}{\degreeCelsius} similar to “medium-temperature baking” and “furnace baking” of cavities is explored \textit{in-situ} by synchrotron X-ray photoelectron spectroscopy (XPS).
Our findings imply that below the critical thickness of the $Nb_2O_5$ layer ($\approx\SI{1}{nm}$) niobium starts to interact actively with surface impurities, such as carbon and phosphorus.
By studying the kinetics of the native oxide reduction, the activation energy and the rate-constant relation have been determined and used for the calculation of the oxygen-concentration depth profiles.
It has been established that the controlled diffusion of oxygen is realized at temperatures \SIrange{200}{300}{\degreeCelsius}, and the native-oxide layer represents an oxygen source, while at \SI{400}{\degreeCelsius} the pentoxide is completely reduced and the doping level is determined by an ambient oxygen partial pressure.
Fluorine (F to Nb atomic ratio is 0.2) after the buffered chemical polishing was found to be incorporated into the surface layer probed by XPS ($\approx\SI{4.6}{nm}$), and its concentration increased during the low-temperature baking (F/Nb=0.35 at \SI{230}{\degreeCelsius}) and depleted at higher temperatures (F/Nb=0.11 at \SI{400}{\degreeCelsius}). 
Thus, the influence of fluorine on the performance of mid-T baked, nitrogen-doped and particularly mild-baked (\SI{120}{\degreeCelsius}/\SI{48}{\hour}) cavities must be considered.
The possible role of fluorine in the educed 
$Nb^{+5}$ → $Nb^{+4}$ reaction under the impact of an X-ray beam at room temperature and during the thermal treatment is also discussed.
The range of temperature and duration parameters of the thermal treatment at which the niobium surface would not be contaminated with impurities is determined.  
	\end{abstract}
}

\begin{document}


\maketitle 


\section{Introduction}
\label{introduction}
The final vacuum thermal treatment (or baking) of superconducting radiofrequency (SRF) niobium cavities is a key technological step in cavity production.  It is used for improving the cavity intrinsic quality factor $Q_0$ \autocite{padamsee2008rf}.
A classic procedure consisting of vacuum annealing of a cavity at \SI{120}{\degreeCelsius} during \SI{48}{\hour} (often called “mild baking” or “low-temperature baking” which is a part of the baseline treatment) has been widely used to mitigate the high-field $Q_0$ decrease (a Q-slope) in $Q_0$($E_{acc}$) dependence up to now \autocite{visentin1999cavity, casalbuoni2005surface}. 
For more than two decades there has not been a commonly accepted explanation of the observed phenomenon on cavities with regard to chemical composition and structure of niobium surface. The change of the near-surface dislocation content \autocite{romanenko2010role}, doping of the niobium surface with vacancy-hydrogen complexes \autocite{visentin2010involvement, romanenko2013effect}, and enrichment \autocite{ma2003angle, delheusy2008x} or depletion \autocite{ciovati2006improved} of oxygen interstitial concentration in the near-surface region caused by dissolution of the native oxide into niobium were proposed to explain the increase of the $Q_0$. 
Later, “nitrogen doping” (“N-doping”) was elaborated \autocite{grassellino2013nitrogen, halbritter1989process} which resulted in a three-times higher $Q_0$ and “an anti-Q-slope” testifying the decrease of surface resistance with $E_{acc}$. It has been shown that such a procedure results in nitrogen-atoms enrichment of the niobium surface layer within the RF penetration depth which decreases the BCS resistance by the decrease of the mean-free path of the electrons in the normal state \autocite{padamsee2008rf}.
The presence of oxygen, nitrogen and carbon interstitials in a metal lattice often is related to the hydrogen trapping phenomenon which further contributes to the reduction of losses by suppressing the formation of niobium hydrides at low temperatures \autocite{pfeiffer1976trapping, trenikhina2015nanostructural}. 
These promising results triggered more research in this direction. In 2017, “nitrogen infusion” treatment consisting in \SI{800}{\degreeCelsius} baking (\SI{3}{\hour}) followed by nitriding at lower temperatures (\SIrange{120}{200}{\degreeCelsius}, \SIrange{24}{48}{\hour}) has shown similar results on cavities \autocite{grassellino2017unprecedented}. This treatment is resembling the standard “mild baking” but the state of the niobium surface subjected to a low-temperature anneal is different.
In the mild baking the surface of niobium is covered with a native oxide after the final chemical etching and air exposure while in the latter, the native oxide is absent since it was dissolved upon a preceding \SI{800}{\degreeCelsius}/\SI{3}{\hour}-step. In the present work it will be shown that the elemental composition of the niobium surface is also different in these recipes. The success of infusion was not always reproducible \autocite{wenskat2020nitrogen}. Moreover, the information on the interaction of nitrogen with niobium at such low temperatures and the resulting surface chemical composition is discrepant \autocite{checchin2018new, semione2019niobium}.
In our previous work the role of nitrogen gas at \SI{120}{\degreeCelsius} in such treatment was investigated with synchrotron X-ray photoelectron spectroscopy (XPS) by studying the niobium surface naturally oxidized after the contact with air \autocite{prudnikava2022systematic}, and no evidence of Nb-N interaction was determined consistent with \autocite{semione2019niobium}.
Recently, so-called “medium-temperature baking” (mid-T baking) and a similar “furnace baking” have raised interest in the SRF community \autocite{posen2020ultralow, ito2021influence}. Both treatments consist of vacuum annealing of a cavity at temperatures \SIrange{200}{450}{\degreeCelsius} for different durations, and in the case of furnace baking the cavity is subsequently air-exposed and high-pressure rinsed (HPRed) prior to RF testing.

The performance of cavities treated at these temperatures have been previously investigated as well \autocite{palmer1987influence, palmer1990oxide}. In particular, dissolution at \SI{300}{\degreeCelsius} of the native oxide grown on \SI{8.60}{\giga\hertz}-cavities previously annealed at \SI{1400}{\degreeCelsius} showed a \SI{20}{\percent}-decrease of the BCS resistance ($R_{BSC}$) and a \SI{40}{\nano\ohm} increase in the residual resistance ($R_{res}$) \autocite{palmer1987influence}.
It was determined that the changes in the BCS resistance were caused by shortening of the electron mean free path by oxygen that had diffused into the metal \autocite{palmer1990oxide}.
The increase of the residual resistance was not exactly identified, while “oxygen-induced surface roughening” was suggested to have an impact. In these experiments the oxide grown in oxygen atmosphere for two days was very thin (\SI{1.3} {\nano\meter}) and was represented by Nb$^{+5}$ and Nb$^{+1}$ chemical states as identified by XPS. 

The procedure of recent experiments on mid-T baking was slightly different and more successful, since it reduced both $R_{BSC}$ and $R_{res}$. In particular, in \autocite{posen2020ultralow} the baking was performed after the standard preparation of a \SI{1.3}{\giga\hertz}-cavity including the electropolishing and HPR.
The best results in this work were obtained at $\approx$\SI{300}{\degreeCelsius}/\SI{2.5}{\hour} according to the cavity cold RF tests conducted immediately after the baking (neither air nor water exposure) which showed $Q_0$-values as high as \num{4e11} at $E_{acc}=\SI{30}{\mega\volt\per\meter}$, $T=\SI{1.4}{\kelvin}$, and \num{5e10} at $E_{acc}=\SI{30}{\mega\volt\per\meter}$, $T=\SI{2}{\kelvin}$. A BCS resistance of \SI{5}{\nano\ohm} and the residual resistance of \SI{0.63}{\nano\ohm} were demonstrated.
SIMS analysis of similarly treated samples showed a sharp drop in concentration of Nb$_2$O$_5$\textsuperscript{-} ions, as well as the increase in nitrogen content as compared to the base-line treated niobium. The reduction of a residual resistance was related to dissolution of the native oxide and diffusion of impurities. Re-oxidation of the cavity in water resulted in increased residual losses. 

In \autocite{ito2021influence} thermal treatments at \SIrange{200}{800}{\degreeCelsius}/\SI{3}{\hour} were investigated but all the cavities were HPRed prior to RF tests which is a peculiarity of the furnace baking. Despite of this, the cavities treated at \SIrange{300}{350}{\degreeCelsius}/\SI{3}{\hour} demonstrated high $Q_0$-values and anti-Q slope, and the highest $Q_0$ of \num{5e10} at \SI{16}{\mega\volt\per\meter} was obtained for \SI{300}{\degreeCelsius}/\SI{3}{\hour}-baked cavities. 
The cavities annealed at \SIrange{600}{800}{\degreeCelsius} did not indicate any increase of efficiency which was related to higher $R_{BCS}$ though the lowest values of $R_{res}$ were calculated.
No characterization of the niobium surface was performed in these works.

The mid-T baking was also tested on cavities at higher temperatures and different thermal pre-treatments on large-grain niobium cavities \autocite{yang2022surface}, with nitrogen-gas addition \autocite{yang2022effective} and also on \SI{650}{\mega\hertz} cavities \autocite{sha2022quality}.
The surface of the Nb samples treated together with cavities was studied primarily by SIMS that allowed determination of the concentration profile in the surface layer characterized mainly by the increase in oxygen-, carbon- and nitrogen-containing species. No change of hydrogen as compared to the baseline treatment was observed \autocite{yang2022surface, yang2022effective}. In \autocite{lechner2021rf} SIMS-depth profiling of the specially prepared samples using the recipes of the furnace baking \autocite{ito2021influence} demonstrated a large increase in oxygen concentration and small quantities of C and N (though larger ones as compared to the electropolished niobium surface).
Using the experimental oxygen-depth profiles and a known theoretical model of diffusion, both the parameters of the oxide-dissolution kinetics and the oxygen-atomic diffusion into niobium were determined. The explored data range covered the temperatures up to \SI{350}{\degreeCelsius} but no limits for validity of the used model were noted (temperature range, duration of the anneal, oxide thicknesses). The authors conclude that the performance of mid-T baked cavity “may be fully explainable in terms of oxygen as the alloying diffusant” and that “interstitial nitrogen may be a secondary effect” \autocite{lechner2021rf}.

Despite some data on the composition of the niobium surface prepared with mid-T baking has been gained, no general conclusion has been made on the optimal treatment parameters of mid-T baking, or whether it is possible to improve the treatment procedure.
The lack of thorough knowledge of the state of niobium surface upon particular chemical and thermal treatment leads to misinterpretation of the results of cavity tests.

In the present work we have chosen the XPS technique to study the mid-T baking and furnace baking processes. XPS allows determination of the binding energies of occupied states in the core-level region and thus identify the chemical state of elements and estimate their concentration in the surface area.
Synchrotron-radiation-excited XPS is a well-controlled highly sensitive technique which allows \textit{in-situ} monitoring of the chemical changes occurring at the niobium surface during the heat treatment. 

XPS has been previously used to study the niobium-oxide dissolution at various conditions \autocite{ma2003angle, ma2004thermal, arfaoui2004model}. B.R. King studied the oxygen-dissolution kinetics of the anodic-oxide films of various thicknesses grown on Nb-Zr(\SI{1}{\percent}) alloy foils \autocite{king1990kinetic}.
In this work the resistive heating was employed which provided very high heating rates (from \SI{423}{\kelvin} to \SI{673}{\kelvin} in less than \SI{7}{\second}) necessary for kinetics measurements. 
Anodized niobium was also studied in \autocite{dacca1998xps} upon heating at temperatures \SIrange{200}{1000}{\degreeCelsius} (\SI{0.7}{\degreeCelsius\per\minute}). 
M. Delheusey explored the oxide reduction on high-purity single-crystal niobium Nb(110) that was preliminary annealed at \SI{2100}{\degreeCelsius} to avoid any foreign impurities \autocite{delheusy2008x}.
The annealing regimes included \SI{145}{\degreeCelsius}/\SI{5}{\hour} and \SI{300}{\degreeCelsius}/\SI{50}{\minute} in UHV. The samples were annealed \textit{in-situ}, but the measurements were performed at room temperature in this work.
The oxide-metal interface of Nb(110) annealed at \SIrange{1300}{1500}{\kelvin} was studied with synchrotron-radiation-excited XPS in \autocite{arfaoui2002tiling, arfaoui2004model}.
The studies more similar to mid-T baking were described in \autocite{ma2003angle, ma2004thermal}. The fine-grain Nb processed with BCP was treated at \SI{250}{\degreeCelsius}/\SI{20}{\hour} \autocite{ma2003angle}, and at \SI{430}{\kelvin}/\SI{30}{\hour} and subsequently at \SI{550}{\degreeCelsius}/\SI{8}{\hour} \autocite{ma2004thermal}. However, all those treatments do not mimic the exact parameters of mid-T baking. 

To complement the aforementioned studies and to clarify the surface state upon the mid-T baking procedure, in the present work we employ a variable-energy synchrotron XPS to study the niobium upon vacuum thermal treatments at temperatures and durations similar to mid-T baking and furnace baking. Detailed information on the chemical state of niobium surface was traced \textit{in-situ}, i.e. before, during, and after the baking, as well as after the air exposure.
In particular, the annealing of fine-grain niobium at \SI{200}{\degreeCelsius}, \SI{230}{\degreeCelsius}, \SI{300}{\degreeCelsius} and \SI{400}{\degreeCelsius} during \SIrange{3}{15}{\hour} has been studied. For the data interpretation the previously established fitting model with the identified chemical shifts of niobium bonded to oxygen and carbon within the Nb 3$d$ core level has been used \autocite{prudnikava2022systematic}. 
The process of the native-oxide dissolution over the heat treatment stages has been studied by monitoring the changes within the Nb 3$d$ and O 1$s$ core levels, which allows estimation of the kinetic parameters of the oxide dissolution and also calculate the oxygen concentration profile in niobium upon respective treatment.
The “behavior” of the surface impurities naturally present at the surface of niobium cavities after the buffered chemical polishing and air exposure, such as carbonaceous layer, phosphorus-containing species, and, most importantly, fluorine with niobium upon these treatments will be discussed. Based on our findings, recommendations on how to diminish or avoid contamination of the niobium surface with external impurities during baking will be given.
The potential future directions of R\&D on niobium-cavity treatments will be proposed.

\section{Materials and Methods}
\label{methods}
Niobium cut-outs with a diameter of \SI{8}{\milli\meter} and a height of \SI{2.8}{\milli\meter} were prepared by electrical discharge machining from the fine-grain niobium manufactured by Tokyo Denkai and used for the EU-XFEL cavities production.
Each sample was treated chemically in a buffered chemical solution ($HF$:$HNO_3$:$H_3PO_4$ = 1:1:2 by volume) for \SI{20}{\minute} at room temperature followed by a thorough washing in a stream of ultra-pure water.
Prior to the experiments, the as-treated niobium cut-outs were kept in air environment (4–8 days) in order to build up a sufficient layer of native oxide on their surface.

\begin{figure*} 
	\includegraphics[width=\linewidth]{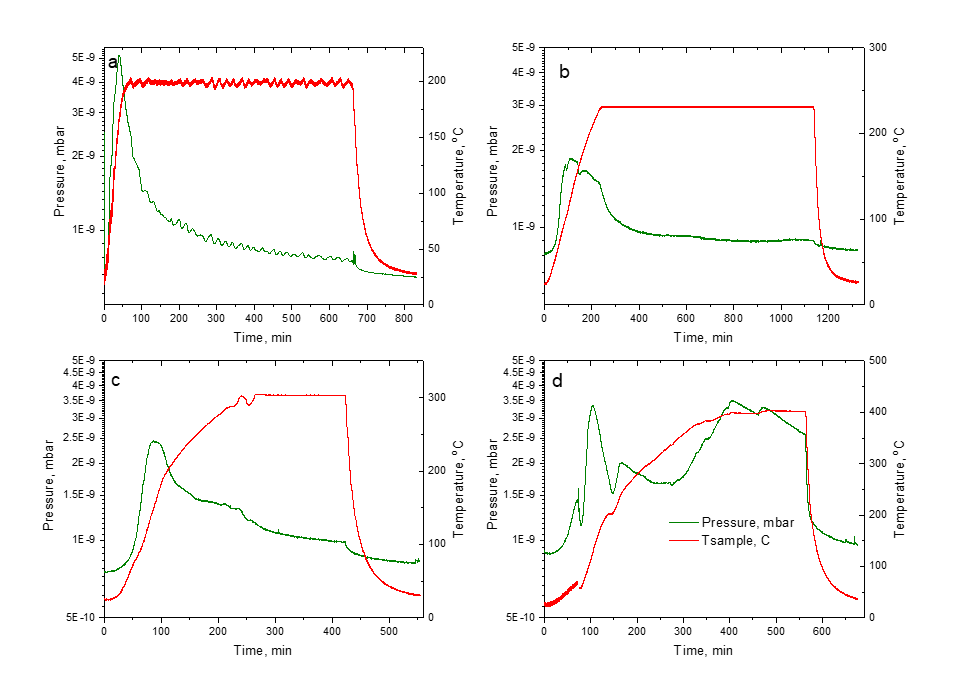}
	\caption{(p, T) vs. t plots during mid-T baking at 200, 230, 300 and 400 °C.}
	\label{fig:PandTvst}
\end{figure*}

Thermal treatments of niobium were performed in the ultra-high vacuum chamber at RGBL beamline at BESSY II (HZB, Berlin) \autocite{fedoseenko2003commissioning}.
The samples were heated by electron bombardment produced by thermal emission from a tungsten filament at the back side of the sample.
The temperature was controlled by a thermocouple attached to a molybdenum sample holder. 
Beforehand the thermocouple was calibrated using a dummy niobium sample with an additional thermocouple welded directly to its surface.
The heat-treatment experiments were performed at temperatures in the range from \SI{200}{\degreeCelsius} to \SI{400}{\degreeCelsius} and base pressure \SIrange{1e-9}{5.2e-9}{\milli\bar}, and the heating rate up to \SI{1}{\degreeCelsius\per\minute} (see Figure \ref{fig:PandTvst}).
It was established that at higher heating rates the base pressure in the chamber could rise up to \SI{4.9e-8}{\milli\bar} due to outgassing from the sample.

In this paper we present the results of baking conducted at \SI{300}{\degreeCelsius} and \SI{400}{\degreeCelsius} for three hours, as well as at \SI{200}{\degreeCelsius} for \SI{11.5}{\hour}, and at \SI{230}{\degreeCelsius} for \SI{15}{\hour}. 

The heat treatment procedure consists of three stages: heating-up, baking and cooling-down.
In the heating-up stage above \SI{200}{\degreeCelsius} the heating rate was kept close to \SI{1}{\degreeCelsius\per\minute} to reproduce the heating rate of the industrial furnaces used for the cavity treatment. 
When the desired temperature was reached, it was maintained constant during the baking stage (for \SI{3}{\hour}, \SI{11.5}{\hour} and \SI{15}{\hour} in different treatments).
When the baking completed, the sample heating was switched off and the sample cooled down naturally (mainly by radiation).

The residual gas analyzer used to measure gas species in the analytic chamber during the heat treatment detected the following molecules: CO, CO$_2$, H$_2$O.

Most of the XPS spectra were collected \textit{in-situ} during baking at photon-beam energies of 900 or \SI{1000}{\electronvolt} and normal-emission geometry at a \SI{55}{\degree} angle between the incident beam and the analyzer aperture. 
In certain cases, for example, to study the surface contaminant species with more precision (to provide higher surface sensitivity) or to perform quantitative analysis, lower photon energies were used. 

The narrow-band high-resolution photoemission-energy distribution curves (EDCs) were collected with the  pass energy of \SI{10}{\electronvolt}. 
The binding-energy (BE) scale of the spectra was calibrated using the Au 4$f_{7/2}$ core-level peak located at \SI{84.0}{\electronvolt} measured at a metal foil.
The EDCs of interest in the present work included Nb 3$d$, O 1$s$, C 1$s$, N 1$s$, P 2$p$, F 1$s$ with the main focus on Nb 3$d$ core-level analysis. 

The spectra were fitted in the CasaXPS software package using the fitting model described in \autocite{prudnikava2022systematic}. Briefly, the background was fitted with the iterated Shirley-type-curve with offsets \autocite{engelhard2020introductory}.
A linear dependence of the binding energy shifts of oxidation states in niobium oxides was considered when fitting the Nb 3$d$ core-level region.
All the spectra were fitted with the metallic Nb\textsuperscript{0} doublet and the doublets corresponding to niobium oxides with the oxidation states ranging from +5 to +1.
Contributions from oxygen interstitials within the niobium crystal lattice were included in the model as Nb\textsuperscript{+0.6} and Nb\textsuperscript{+0.4}.
The lineshapes of Nb 3$d$ core-level peak features were chosen symmetric or asymmetric depending on the conductivity type of the respective compound in a bulk state. 
The lineshapes corresponding to Nb$^{+5}$, Nb$^{+4}$ and Nb$^{+3}$ oxides collected during baking had larger full widths at half maximum (FWHM) as compared to those at room temperature which is related to the oxide reduction process.

The information depth, $S$, was estimated according to \autocite{powell2010progress}:

\begin{equation}
	S = \lambda\cos\alpha\ln\frac{1}{1-(P/100)} 
	\label{eq:info_depth}
\end{equation}
where $\lambda$ is the inelastic mean free path (IMFP) for the photoelectrons in the sample, $\alpha$ is the angle of emission with respect to the surface normal, and $P$ is a specified percentage of the detected signal. 
The IMFPs of niobium and its compounds were calculated using a universal predictive equation for the IMFPs of X-ray photoelectrons and Auger electrons (G1 equation) described in \autocite{gries1996universal}.
The obtained values of the \SI{90}{\percent}-information depth in our experimental setup at $h\nu$=\SI{1000}{\electronvolt} (\SI{400}{\electronvolt}) are \SI{4.64}{\nano\meter} (\SI{2.25}{\nano\meter}) for Nb$_2$O$_5$ and \SI{3.80}{\nano\meter} (\SI{1.87}{\nano\meter}) for pure Nb, respectively.

The samples were additionally characterized by SEM (Zeiss Merlin) equipped with an Energy-Dispersive Analyzer (Ultim Extreme, Oxford Instruments). 
Furthermore, the X-ray diffraction measurements were performed with the Bruker D8 Advance system employing a Cu K$\alpha$ anode both in Bragg–Brentano focusing geometry ($2\Theta/\omega$ scans) as well as at various angles $\omega$ of an incident photon beam.

\section{Results}
\label{results}
\subsection{SEM/XRD}
\begin{figure} 
	\includegraphics[width=\linewidth]{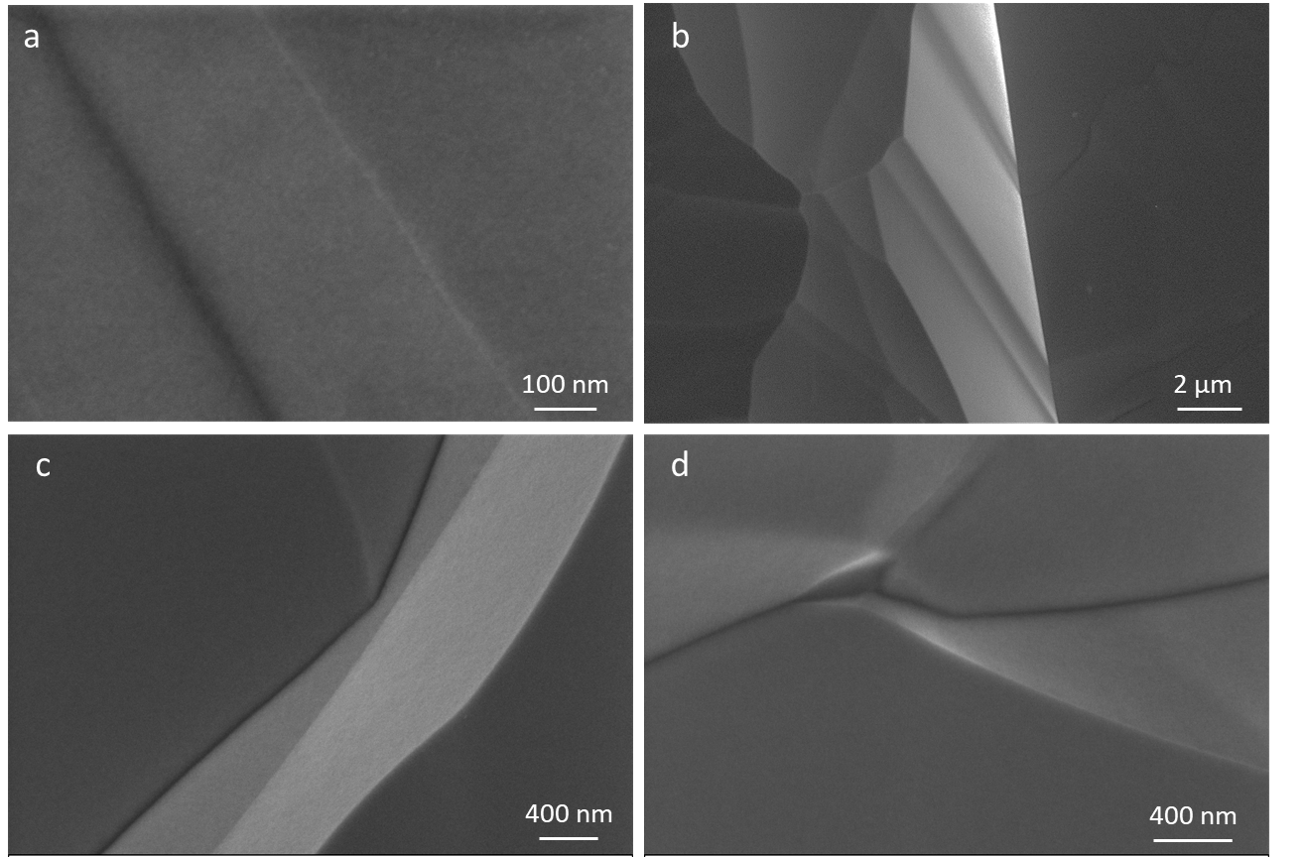}
	\caption{SEM images of the annealed niobium surface: (a) 200°C/11.5h, (b) 230°C/15h, (c) 300°C/3h, (d) 400°C/3h.}
	\label{fig:SEM_annealed}
\end{figure}

Inspection of the thermally treated niobium samples with SEM in secondary electron mode did not reveal any precipitates of new phases compared to the untreated surface (see Figure \ref{fig:SEM_annealed}).
A few organic particles from the contact with environment as well as calcium oxide that were also traced by XPS were found.
According to the EDX elemental mapping, some areas of the initial and the annealed at \SIrange{200}{230}{\degreeCelsius} samples contained trace amount of phosphorous (P/Nb=0.01, $E_{acc}$=\SIrange{3}{10}{\kilo\volt}) and fluorine (F/Nb=0.006-0.1), while in \SI{400}{\degreeCelsius}-sample no fluorine could be detected by EDX.
No P-containing particles were found.
The X-ray diffraction analysis did not reveal any reflexes other than niobium (not shown).

\subsection{Initial Surface State: Niobium after BCP}
\begin{figure} 
	\includegraphics[width=\linewidth]{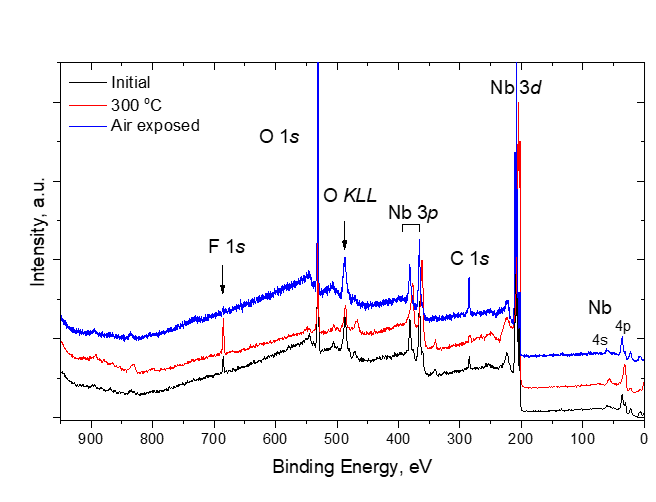}
	\caption{Survey spectra ($h\nu$=\SI{1000}{\electronvolt}) collected from the initial niobium (black), annealed at 300 °C (red), and after the air exposure (blue).}
	\label{fig:XPSSurveySpectra300C}
\end{figure}

The survey low-resolution EDCs (pass energy \SI{50}{\electronvolt}, $h\nu$=\SI{1000}{\electronvolt}) of the niobium samples after BCP demonstrate classical features associated with niobium, oxygen, carbon, and fluorine (see Figure \ref{fig:XPSSurveySpectra300C}). 

\begin{figure*} 
	\includegraphics[width=\linewidth]{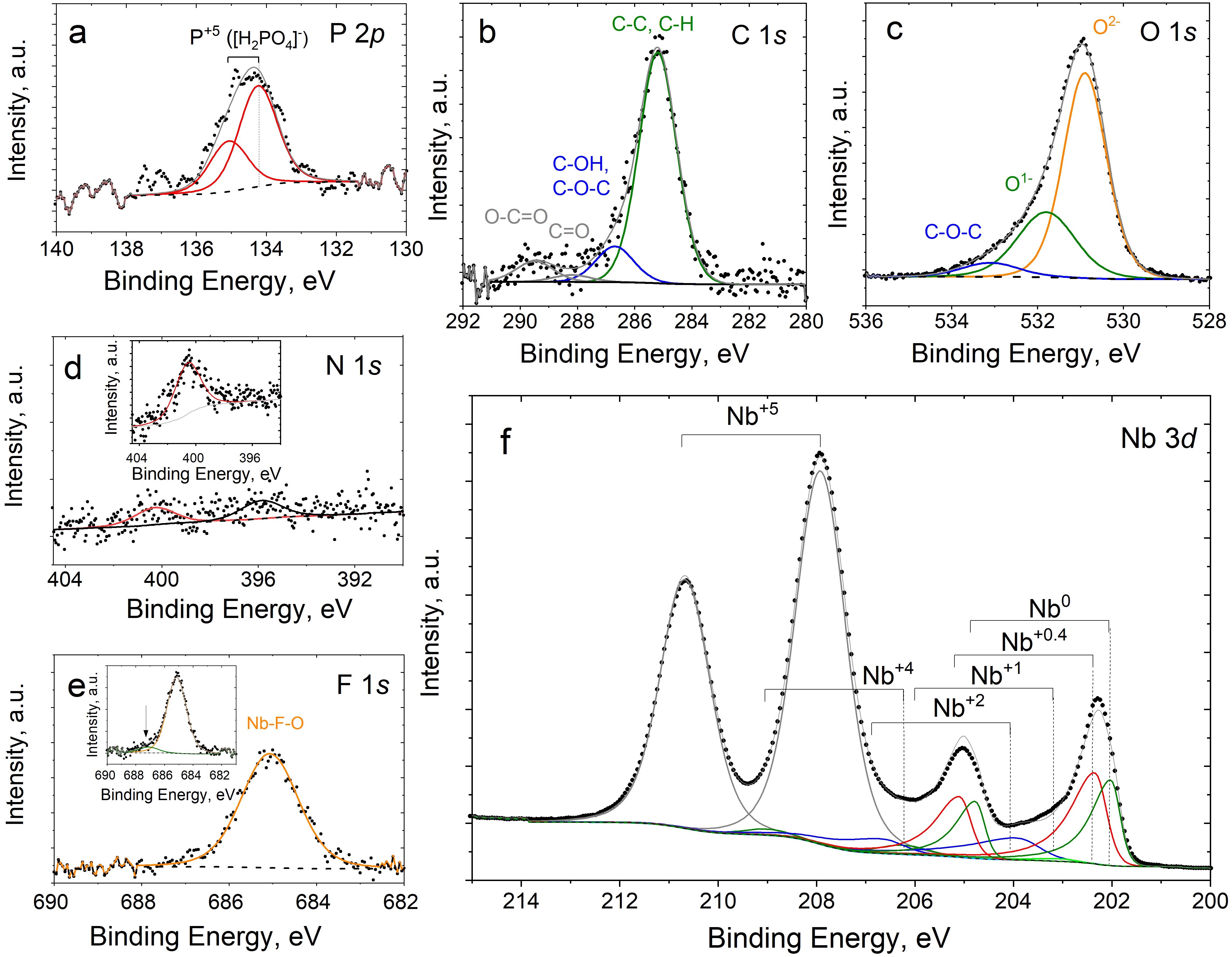}
	\caption{Typical high-resolution EDCs of P 2p (a), C 1s (b), O 1s (c), N 1s (d), F 1s (e), and Nb 3d (f) collected at $h\nu$=\SI{1000}{\electronvolt} for the initial niobium surface. Inset (d): The EDC of N 1s collected at $h\nu$=\SI{480}{\electronvolt}. Inset (e): The EDC of F 1s after the prolonged XPS measurements.}
	\label{fig:XPS_typical}
\end{figure*}

The typical high-resolution EDCs ($h\nu$=\SI{1000}{\electronvolt}) for the regions of interest of the niobium sample after the chemical etching are shown in Figure \ref{fig:XPS_typical}(a-f).

A weak peak P 2$p_{3/2}$  at BE=\SI{134.25}{\electronvolt} (Figure \ref{fig:XPS_typical}(a)) was detected which according to \autocite{rumble1992nist, siow2018xps} may correspond to phosphate (PO$_4$)\textsuperscript{–} or metaphosphate (PO$_3$)\textsuperscript{–}.
The roughly estimated position of P 2$s$ peak in the survey EDCs (not visible in Figure \ref{fig:XPSSurveySpectra300C}) at \SI{191.7}{\electronvolt} according to \autocite{wagner2003nist} may be a characteristic of (PO$_3$)\textsuperscript{–}.
According to \autocite{rumble1992nist}, the BE of P 2$p_{3/2}$ is close to valence five, so the phosphorus is most likely in P\textsuperscript{+5}-state. 
From these data it is hard to distinguish whether it is a salt, a functional group or an anion.
Several compounds originating from the contact of niobium or its oxides with the phosphoric acid may fulfil these criteria.
Among the most probable ones is (H$_2$PO$_4$)\textsuperscript{–} anion which may be incorporated into the oxide films typically anodic ones \autocite{jouve1986x, bokii1989x}, or being adsorbed on the surface \autocite{francisco2004surface, pavan2005adsorption}.
Niobium phosphate NbOPO$_4$ was also reported \autocite{okazaki1993surface}.
Since the intensity of P 2$p$ peak is very weak, and it rises as the energy of the primary X-ray beam decreases, we deem that it may rather correspond to the ionic species adsorbed to the niobium surface rather than being incorporated into the niobium-oxide film.

\begin{figure} 
	\includegraphics[width=\linewidth]{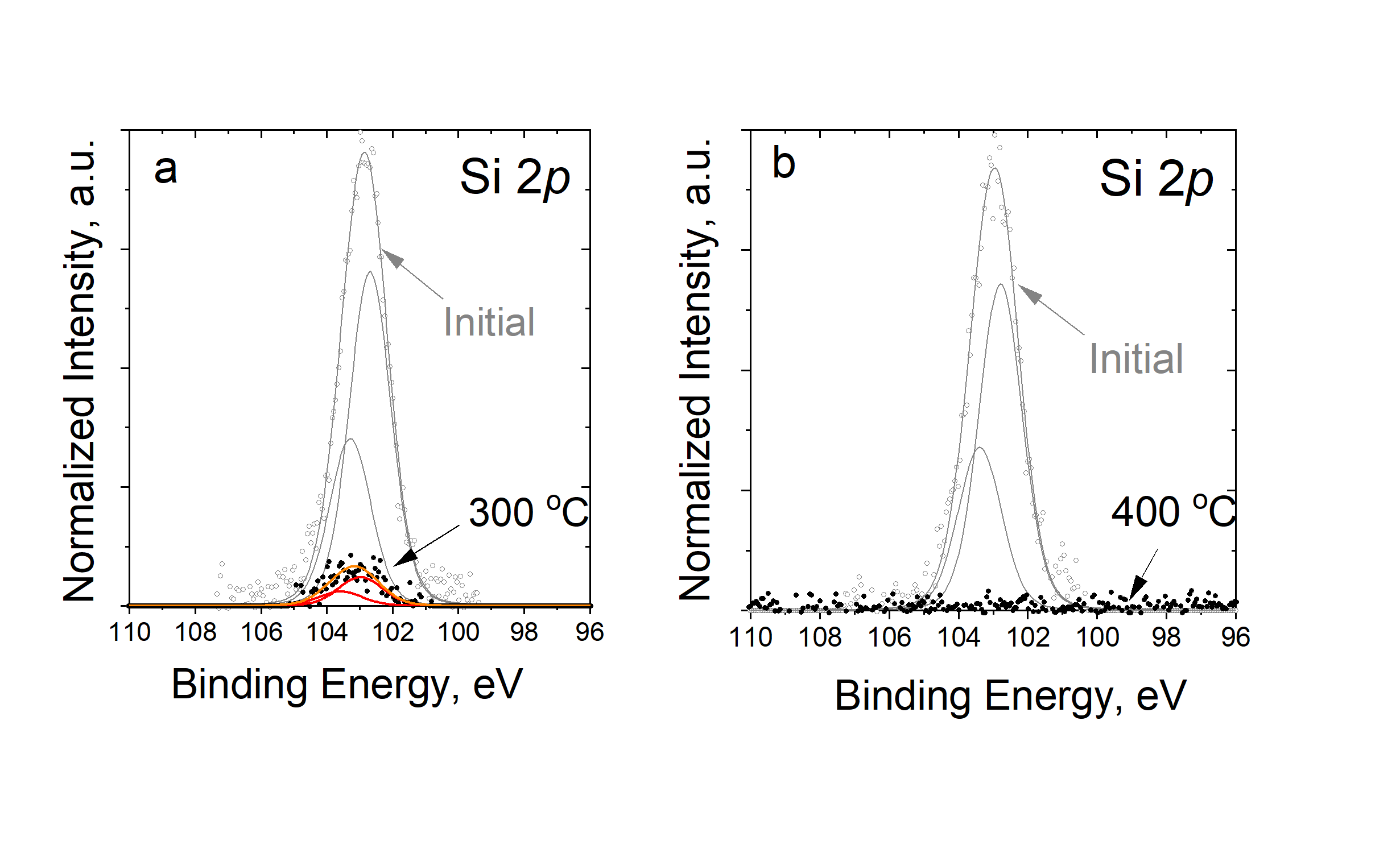}
	\caption{The EDC of Si 2p ($h\nu$=\SI{480}{\electronvolt}) before and after the annealing of Nb at \SI{300}{\degreeCelsius} (a) and \SI{400}{\degreeCelsius} (b).}
	\label{fig:XPS_Si2p_detailed}
\end{figure}

Trace amounts of silicon (atomic ratio Si/Nb = 0.01-0.02) in a state of organic silicon (BE=\SI{102.3}{\electronvolt}) were also detected (see Figure \ref{fig:XPS_Si2p_detailed}).

The C 1$s$ core-level is represented by carbonaceous adsorbates, a contamination layer commonly found at the surfaces of solids after the air exposure (Figure \ref{fig:XPS_typical}(b)).
It was fitted with four contributions.
The most intense is located at \SIrange{285.1}{285.2}{\electronvolt} and interpreted as non-graphitic C–C and C–H bonds.
The peaks of lower intensity may be assigned to C-OH, C-O-C (\SI{286.7}{\electronvolt}), C=O (\SI{288.2}{\electronvolt}), and O-C=O (\SI{289.5}{\electronvolt}).

O 1$s$ is fitted with three peaks located at \SI{530.9}{\electronvolt}, \SI{531.8}{\electronvolt} and \SI{533.1}{\electronvolt} (Figure \ref{fig:XPS_typical}(c)).
The first is the most intense and is usually interpreted as O\textsuperscript{-2} ions in niobium oxides while the second peak is assigned to low-coordination-number O\textsuperscript{-1} ions that are compensating for deficiencies in the oxide subsurface \autocite{dupin2000systematic}.
The feature at around \SI{533.1}{\electronvolt} is associated with weakly adsorbed species such as aliphatic C-O-C, C-OH, etc. and possibly (H$_2$PO$_4$)\textsuperscript{–} \autocite{beamson1992high}. 

The intensity of N 1$s$ was very weak at $h\nu$=\SI{1000}{\electronvolt} (Figure \ref{fig:XPS_typical}(d)).
Two broad peaks could presumably be fitted at around \SI{400}{\electronvolt} and \SI{395.7}{\electronvolt} that could be related to nitrogen containing species adsorbed at the surface inhomogeneities after exposure to the ambient environment and to nitrogen in Nb lattice, respectively.
For more information, the EDC of N 1$s$ taken at $h\nu$=\SI{480}{\electronvolt} is presented in which only a peak at \SI{400.8}{\electronvolt} from adsorbates is present (Figure \ref{fig:XPS_typical}(d), inset).
At this photon energy the probing depth is much smaller ($\approx$\SI{2.25}{\nano\meter}) so nitrogen in niobium might not be traced. 

F 1$s$ is represented by a single peak located at BE=\SI{685.15}{\electronvolt} confirming the chemical bonding of fluorine with niobium (Figure \ref{fig:XPS_typical}(e)) \autocite{luo2008hydrogen, mandula2017electrochemical}.
A hardly noticeable shoulder at \SIrange{686.8}{687.0}{\electronvolt} (indicated with an arrow in Figure \ref{fig:XPS_typical}(e), inset) appeared during the measurements at room temperature (which will be discussed later.)

The Nb 3$d$ spectrum ($h\nu$=\SI{1000}{\electronvolt}) with the fitted components typical for all niobium samples treated with BCP is presented in Figure \ref{fig:XPS_typical}(f).
Peak maxima of the components, as well as the respective FWHMs estimated by the procedure described in the previous section are summarized in Table \ref{tab:Nb_components_BE}.
The relative percentage areas of the fitted peaks are summarized in Table \ref{tab:XPS_all_components}.

\begin{table*} 
	\centering
    \caption{Binding energy shifts, $\Delta$BE, with respect to Nb\textsuperscript{0} (BE=202.00±0.01), and the full width at half-maximums (FWHMs) of the fitted components of the Nb 3$d_{5/2}$ core-level obtained for the initial niobium samples.}
    \small 
	\begin{tabular}{c c c c c c c c c}
		\toprule
		Oxidation state	& Nb\textsuperscript{+5}	& Nb\textsuperscript{+4}	& Nb\textsuperscript{+3}	& Nb\textsuperscript{+2}	& Nb\textsuperscript{+1} 	& Nb\textsuperscript{+0.6}	& Nb\textsuperscript{+0.4}	& Nb\textsuperscript{0} \\
		\midrule
		BE shift, eV	& 5.95±0.05 & 4.14±0.07 & 3.08±0.02 & 1.91±0.01 & 1.02±0.01 & 0.51±0.03 & 0.31±0.02 & 0\\
        FWHM, eV & 1.18±0.07 & 1.16±0.00 & 1.07±0.00 & 1.13±0.03 & 0.75±0.00 & 0.68±0.04 & 0.54±0.03 & 0.47±0.00\\
		\bottomrule
	\end{tabular}
 	
	\label{tab:Nb_components_BE}
\end{table*}

\begin{table*}
    \centering
    \caption{Relative area of the fitted components of the Nb 3d core-level ($h\nu$=\SI{1000}{\electronvolt}) of niobium before and after the respective anneal.}
    \notsotiny
    \begin{tabular}{cccccccccccc}
        \toprule
        Treatment & Nb\textsuperscript{+5} & Nb\textsuperscript{+4} & Nb\textsuperscript{+3} & Nb\textsuperscript{+2} & Nb\textsuperscript{+1} & Nb\textsuperscript{+0.6} & Nb\textsuperscript{+0.4} & Nb\textsuperscript{0} & Nb-C & Nb-O-F & Nb-O-F (II) \\
        \midrule
        
        \multicolumn{12}{c}{\SI{200}{\degreeCelsius} anneal} \\
        Initial	& 67.36±0.36 & 1.61±0.30 & 0 & 6.61±0.47 & 0.48±0.26 & 0.01±0.01 & 13.58±0.39 & 10.35±0.44 & 0 & 0 & 0\\
        3h & 51.21±0.73 & 12.70±0.60 & 0 & 7.25±0.49 & 0.23±0.45 & 1.28±0.37 & 14.23±0.49 & 13.11±0.59 & 0 & 0 & 0\\ 
        11.5h& 43.52±0.61 & 16.74±0.30 & 0.53±0.51 & 9.85±0.49 & 0.31±0.36 & 0.28±0.23 & 12.99±0.27 & 15.77±0.30 & 0 & 0 & 0\\
        \midrule
        
        \multicolumn{12}{c}{\SI{230}{\degreeCelsius} anneal*} \\
        Initial & 68.67±0.53 & 0.94±0.46 & 0 & 6.23±0.67 & 2.11±0.09 & 0 & 12.70±0.87 & 9.35±0.90 & 0 & 0 & 0\\
        3h & 30.76±0.76 & 13.60±0.48 & 13.32±0.88 & 7.52±0.71 & 1.64±0.28 & 0 & 14.32±0.67 & 15.50±0.82 & 0 & 0 & 3.34±0.36\\
        15h & 16.88±0.47 & 11.54±0.30 & 14.17±0.20 & 16.37±0.60 & 2.32±0.14 & 0 & 17.27±0.42 & 15.66±0.49 & 1.72±0.17 & 0 & 4.06±0.16\\
        Air & 77.14±0.31 & 3.03±0.30 & 0 & 3.19±0.22 & 1.36±0.14 & 2.30±0.08 & 7.60±0.09 & 3.35±0.11 & 1.45±0.12 & 0 & 0.57±0.16\\
        \midrule
        \multicolumn{12}{c}{\SI{300}{\degreeCelsius} anneal} \\
        Initial & 65.94±0.76 & 1.11±0.67 & 0 & 6.07±0.76 & 3.33±0.61 & 0.01±0.01 & 15.94±0.76 & 7.61±0.58& 0 & 0 & 0\\
        3h & 5.42±0.60 & 1.93±0.54 & 1.39±0.97 & 18.88±0.67 & 13.85±0.69 & 4.70±0.65 & 22.28±0.73 & 17.39±1.03 & 6.82±0.52 & 2.71±0.54 & 4.62±0.45\\
        Air & 71.18±0.37 & 6.26±0.34 & 0±0 & 3.67±0.26 & 0.80±0.23 & 1.87±0.15 & 10.70±0.13 & 1.73±0.12 & 2.09±0.05 & 0.57±0.02 & 1.13±0.17\\  
        \midrule
        \multicolumn{12}{c}{\SI{400}{\degreeCelsius} anneal} \\
        Initial & 75.67±0.43 & 0.67±0.38 & 0 & 4.89±0.30 & 0.09±0.17 & 0 & 14.58±0.30 & 4.10±0.33 & 0 & 0 & 0\\
        3h & 0 & 0 & 0 & 4.89±0.31 & 9.76±0.42 & 16.21±0.22 & 18.53±0.41 & 34.11±0.45 & 11.57±0.40 & 1.71±0.27 & 3.21±0.37\\
        Air & 59.49±0.23 & 3.64±0.25 & 0±0 & 12.34±0.41 & 0.98±0.16 & 0.25±0.18 & 10.12±0.26 & 6.95±0.14 & 6.24±0.12 & 0 & 0\\    
        \bottomrule
        \multicolumn{5}{l}{*\footnotesize{The Nb 3d core level was measured at $h\nu= \SI{900}{\electronvolt}$.}}
    \end{tabular}    
    \label{tab:XPS_all_components}
\end{table*}

The EDC of Nb 3$d$ collected from the initial niobium was dominated by the Nb\textsuperscript{+5} state (Nb$_2$O$_5$) with minor contributions from lower oxides.
At that stage it has been assumed that fluorine may be distributed within the native oxide layer as Nb$_2$O$_{5-x}$F$_x$ or/and located under the pentoxide.

From the data of Table \ref{tab:XPS_all_components} one can approximately estimate the atomic concentration of oxygen ions found in the state of niobium oxides (Nb\textsuperscript{+5}, Nb\textsuperscript{+4}, Nb\textsuperscript{+3}, Nb\textsuperscript{+2}) and interstitials (the remaining states) within the XPS-information depth.
For calculation, the number of associated oxygen ions in each oxidation state per Nb-atom was considered as well as the Nb\textsuperscript{+5} peak area was corrected (scaled by 0.5) as Nb$_2$O$_5$ contains two atoms of niobium.
The average values obtained for BCP-ed niobium are \SI{58.6}{\atpercent} and \SI{5.3}{\atpercent} for niobium oxides and interstitials, respectively.
These values have been used for the calculation of oxygen-concentration depth profiles.

Upon analysis of the room-temperature Nb 3$d$ core-levels of the initial niobium samples collected from the same location it was revealed that the Nb\textsuperscript{+4} state was initially absent and it appeared subsequently while other core-levels were recorded.
At that, a hardly pronounced shoulder within F 1$s$ at \SI{686.8}{\electronvolt} appeared which is interpreted as C-F bond \autocite{nikolenko2002xps}.
Thus, the formation of Nb\textsuperscript{+4} state at room temperature is most likely associated with the formation of vacancies in the pentoxide due to F-ions transition to an adsorbed state on the sample surface and interaction with carbonaceous layer at the niobium surface.

\subsection{Heat-Treated Niobium: Nd 3d Core-Level.}
\begin{figure*} 
	\includegraphics[width=\linewidth]{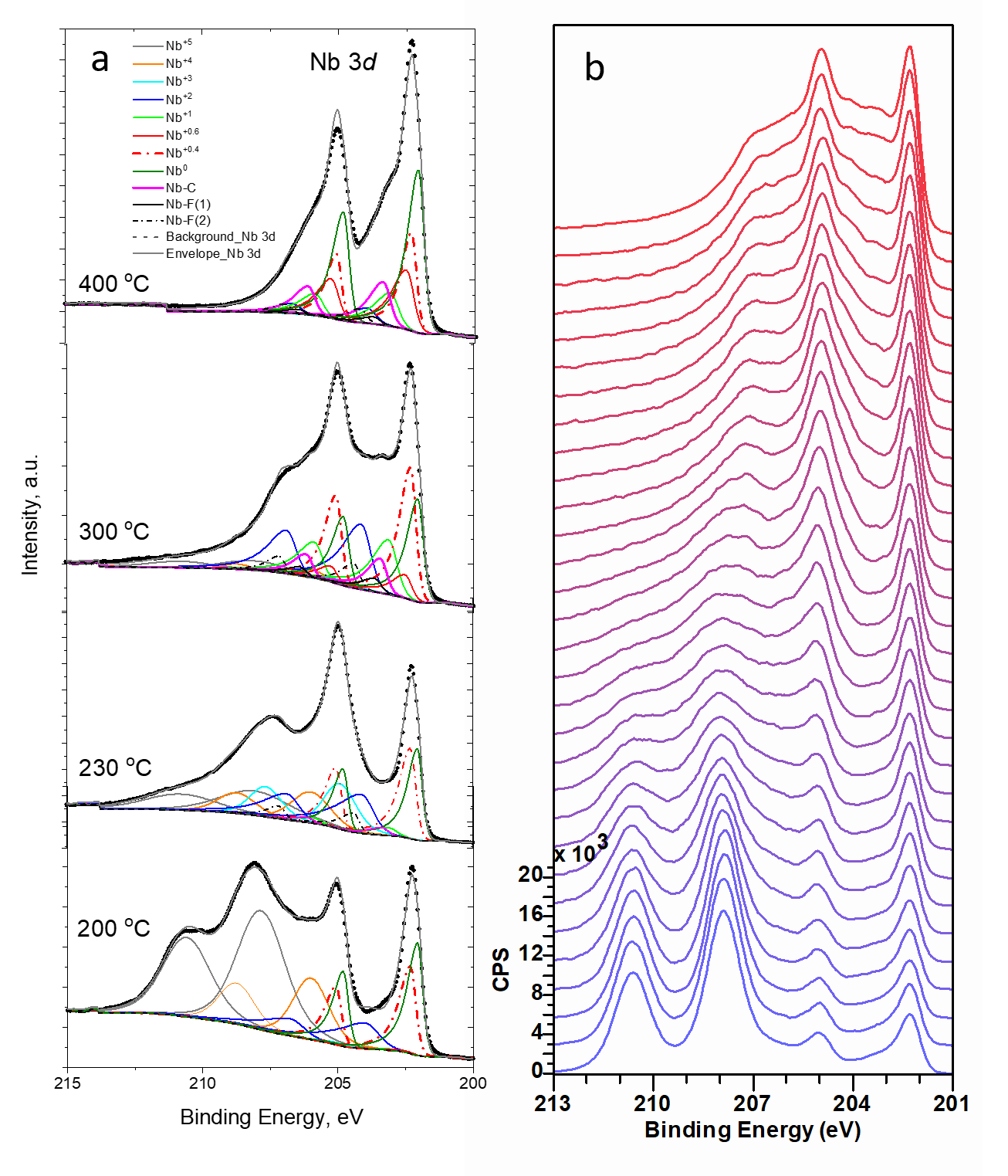}
	\caption{(a) The fitted EDCs of the Nb 3d region ($h\nu$=\SI{1000}{\electronvolt}) of niobium after the respective thermal anneal at 200 °C/11.5 h, 230 °C/15 h, 300 °C/3 h, 400 °C/3 h. (b) Evolution of high-resolution EDCs of the Nb 3d region during the 300 °C anneal starting from the room temperature. The time interval between the neighbouring spectra is 20 min. The spectra are unaltered experimental EDCs with a subtracted Shirley background and non-normalized intensity.}
	\label{fig:XPS_Nb3d}
\end{figure*}

The Nb 3$d$ spectra with the fitted components of the samples baked at 200°C/11.5h, 230°C/15h, 300°C/3h, and 400°C/3h are presented in Figure \ref{fig:XPS_Nb3d}(a).
As an example, the evolution of Nb 3$d$ region measured \textit{in-situ} during baking at \SI{300}{\degreeCelsius} is demonstrated in Figure \ref{fig:XPS_Nb3d}(b).

The FWHMs of the Nb 3$d$ components corresponding to niobium of higher oxidation states (Nb\textsuperscript{+5} and Nb\textsuperscript{+4}) increased during baking.
For example, the FWHM of Nb\textsuperscript{+5} increased from \SI{1.20}{\electronvolt} to \SI{1.84}{\electronvolt} after \SI{3}{\hour} and to \SI{2.10}{\electronvolt} after \SI{11.5}{\hour} of baking at \SI{200}{\degreeCelsius} ($h\nu=\SI{1000}{\electronvolt}$).

For \SI{230}{\degreeCelsius}, the FWHM of Nb\textsuperscript{+5} changed from \SI{1.14}{\electronvolt} to \SI{2.90}{\electronvolt} (\SI{3}{\hour}) and to \SI{3.1}{\electronvolt} (\SI{15}{\hour}) ($h\nu=\SI{900}{\electronvolt}$).
Similar tendency was observed for the FWHM of Nb\textsuperscript{+4}.
On the other hand, the FWHM of the component related to a newly created bond, like for example Nb\textsuperscript{+3}, had a larger value ($\approx$\SIrange{1.4}{1.5}{\electronvolt}) as compared to the average FWHMs of the fitted components within the Nb 3$d$ core-level measured at room-temperature (see Table \ref{tab:Nb_components_BE}).
The broadening of the spectral lines can be related to the defect structure of the crystal lattice and intermediate valence states \autocite{gopel1984surface}.

The relative percentage areas of the fitted peaks after \SI{3}{\hour} anneal are summarized in Table \ref{tab:XPS_all_components}. 
The data for the 200°C- and 230°C-samples that were annealed for longer time are also presented.

During baking at \SI{200}{\degreeCelsius} changes occurred in Nb-O system within the high oxidation states, Nb\textsuperscript{+5} and Nb\textsuperscript{+4}.
Particularly, the relative area of Nb\textsuperscript{+5} decreased from 67\% to 51\% after \SI{3}{\hour} of baking and to 43\% after \SI{11.5}{\hour}.
At that, the relative area of Nb\textsuperscript{+4} increased from 1.6 to 12.7\% after \SI{3}{\hour} and to 16.7\%  after \SI{11.5}{\hour}.
After \SI{3}{\hour} baking, the contribution of Nb\textsuperscript{+2} and Nb\textsuperscript{0} increased insignificantly.
After \SI{11.5}{\hour}, the surface was represented mainly by Nb\textsuperscript{+5} with nearly equal contributions of Nb\textsuperscript{+4}, Nb\textsuperscript{+0.4}, Nb\textsuperscript{0} (within 13–17\%) and slightly less Nb\textsuperscript{+2} (9.8\%) (Figure \ref{fig:XPS_Nb3d}(a, \SI{200}{\degreeCelsius}).

At \SI{230}{\degreeCelsius} the changes in Nb-O system become more evident.
The percentage of Nb\textsuperscript{+5} drops from 69\% to 31\% in \SI{3}{\hour}, and to 17\%  in \SI{15}{\hour}.
The percentage of Nb\textsuperscript{+4} is similar to \SI{200}{\degreeCelsius}-sample, and maintained at 11–13\% during baking.
Additionally, Nb\textsuperscript{+3} bond appears and remains at 13–14\% after both \SI{3}{\hour}- and \SI{15}{\hour}-baking time.
The percentage of Nb\textsuperscript{+2} after \SI{3}{\hour} was comparable (7.5\%) to \SI{200}{\degreeCelsius}/\SI{3}{\hour}, but finally increased to 16\%.
The content of Nb\textsuperscript{+1} and Nb\textsuperscript{+0.6} was small or close to zero in the both samples at various duration of baking.
The quantity of Nb\textsuperscript{+0.4} changed insignificantly with respect to the initial values of $\approx$13\% in both cases (see Table \ref{tab:XPS_all_components}).
Thus, after \SI{15}{\hour} at \SI{230}{\degreeCelsius} the surface is composed almost equally of Nb\textsuperscript{+5}, Nb\textsuperscript{+3}, Nb\textsuperscript{+2}, Nb\textsuperscript{+0.4}, Nb\textsuperscript{0} and slightly less Nb\textsuperscript{+4} (Figure \ref{fig:XPS_Nb3d}(a, \SI{230}{\degreeCelsius})).

For the niobium annealed at \SI{300}{\degreeCelsius}/\SI{3}{\hour}, the Nb\textsuperscript{+5}, Nb\textsuperscript{+4}, Nb\textsuperscript{+3} chemical states almost completely reduced (Figure \ref{fig:XPS_Nb3d}(b)), and the percentage of the remaining chemical states was distributed in the descending order as Nb\textsuperscript{+0.4}(22\%), Nb\textsuperscript{+2}(19\%), Nb\textsuperscript{0}(17\%), Nb\textsuperscript{+1}(14\%) and Nb\textsuperscript{+0.6} (5\%) 
(Figure \ref{fig:XPS_Nb3d}(a, \SI{300}{\degreeCelsius})). 

At \SI{400}{\degreeCelsius}, the oxides Nb\textsuperscript{+5}, Nb\textsuperscript{+4} and Nb\textsuperscript{+3} vanished, and the percentage of chemical states within the Nb 3$d$ core-level was dominated by Nb\textsuperscript{+0.4}(19\%) and Nb\textsuperscript{+0.6} (16\%) followed by Nb\textsuperscript{+1} (10\%) and Nb\textsuperscript{+2}(5\%) (Figure \ref{fig:XPS_Nb3d}(a, \SI{400}{\degreeCelsius})).
The maximal contribution of metallic Nb\textsuperscript{0} was retained within the information depth as compared to the other samples.
This might be related to relatively high diffusion rate of oxygen interstices in Nb-lattice.

Analysis of the data presented in Table \ref{tab:XPS_all_components} shows, that in the initial niobium samples the area of Nb\textsuperscript{+0.6} peak component is approximately zero, and it is getting pronounced upon baking.
Thus, it exists only in the niobium subjected to the thermal treatment, and is the highest for the \SI{400}{\degreeCelsius}-baked Nb.
It was reported that oxidation of niobium at \SIrange{400}{500}{\degreeCelsius} proceeds via creation of Nb$_6$O phase \autocite{chang1969phase} which BE is close to Nb\textsuperscript{+0.4} state.
Alternatively, it can also be assumed that Nb\textsuperscript{+0.6} may refer to a niobium bond with an element other than oxygen (for example, fluorine) which is created during anneal.

The calculated atomic concentrations of oxygen ions found in the state of niobium oxides and interstitials for the baked niobium are presented in Table \ref{tab:Oxygen_components}.
Thus, the number of interstitials in the vicinity of oxide layer approximately doubled at \SIrange{300}{400}{\degreeCelsius} as compared to the initial niobium.
For the prolonged treatments at \SIrange{200}{230}{\degreeCelsius}, these numbers are comparable to the initial ones.

\begin{table*} 
	\centering
    \caption{The concentration of oxygen, at\%, in niobium oxides and interstitial phases 
    in the initial and the annealed Nb samples calculated using the fitted components of Nb 3d core level measured at $h\nu$=\SI{1000}{\electronvolt}. 
    }
	\begin{tabular}{c c c c c}
		\toprule
        \multirow{2}{*}{Treatment} & \multicolumn{2}{c}{Oxygen in NbO$_x$} & \multicolumn{2}{c}{Oxygen in interstitials} \\
                & Initial & Baked & Initial & Baked     \\
        \midrule
        200°C/11.5h & 58.64±1.41 & 55.71±1.54 & 4.27±0.17 & 3.52±0.14 \\
        230°C/15h* & 58.86±2.43 & 47.19±1.00 & 5.19±0.25 & 4.80±0.11 \\
        300°C/3h & 57.50±3.48 & 24.52±1.50 & 6.75±0.50 & 11.62±0.54 \\
        400°C/3h & 61.86±1.49 & 4.66±0.22 & 4.54±0.19 & 11.86±0.37 \\        
		\bottomrule
        \multicolumn{5}{l}{*\footnotesize{The Nb 3d core level was measured at $h\nu= \SI{900}{\electronvolt}$.}}
	\end{tabular} 	
	\label{tab:Oxygen_components}
\end{table*}

\begin{figure*} 
	\includegraphics[width=\linewidth]{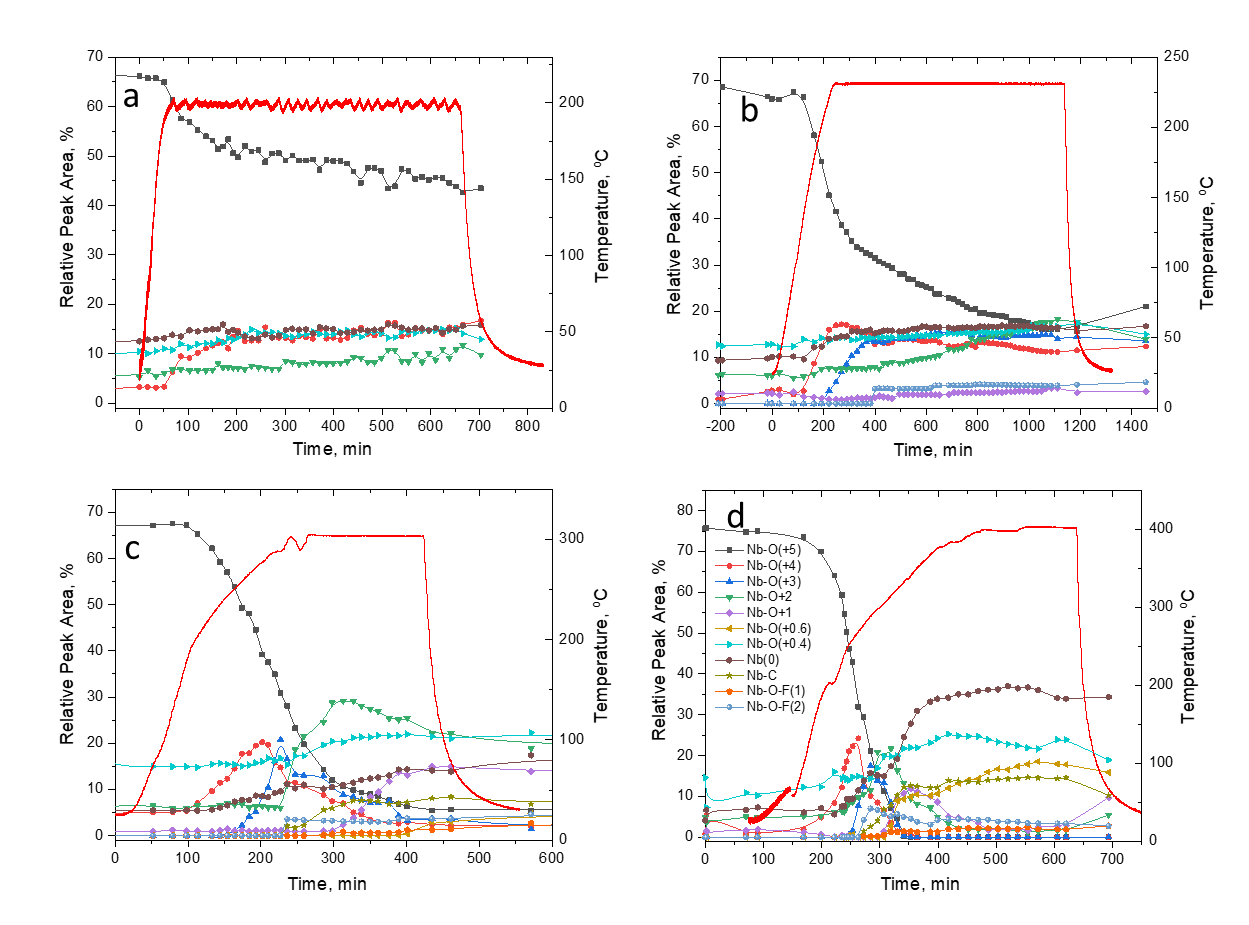}
	\caption{The relative peak areas  of the fitted components during the respective thermal anneal: (a) 200°C/11.5h, $h\nu$=\SI{1000}{\electronvolt}; (b) 230°C/15h, $h\nu$=\SI{900}{\electronvolt}; (c) 300°C/3h, $h\nu$=\SI{900}{\electronvolt}; (d) 400°C/3h, $h\nu$=\SI{900}{\electronvolt}.}
	\label{fig:Components_area}
\end{figure*}

The changes of the relative percentage areas of the fitted peaks within the Nb 3$d$ core-level during these thermal treatments are demonstrated in Figure \ref{fig:Components_area}(a-d).

At \SI{200}{\degreeCelsius} the conversion of Nb\textsuperscript{+5} to Nb\textsuperscript{+4} occurs synchronously and gradually (Figure \ref{fig:Components_area}(a)); Nb\textsuperscript{+2} and Nb\textsuperscript{+0.4} rise slowly; Nb\textsuperscript{+3}, Nb\textsuperscript{+1} and Nb\textsuperscript{+0.6} remain close to zero.

At \SI{230}{\degreeCelsius} reduction of Nb\textsuperscript{+5}  proceeds much faster and a maximum of Nb\textsuperscript{+4} percentage (17\%) was observed at the time when the area of Nb\textsuperscript{+5} was at about 40–45\% (Figure \ref{fig:Components_area}(b)).
Furthermore, at that moment the formation of the Nb\textsuperscript{+3} bonds starts.
The percentages of Nb\textsuperscript{+4}, Nb\textsuperscript{+3}, Nb\textsuperscript{+2}, Nb\textsuperscript{+0.4}, Nb\textsuperscript{0} did not change significantly during the whole duration of baking.
Approximately at the same time a new bond is formed which was assigned to Nb-F bond (associated with the changes within F 1$s$, which will be described later).

At \SI{300}{\degreeCelsius}, the Nb\textsuperscript{+4} and Nb\textsuperscript{+3} states appeared during the heating-up stage and vanished by the end of baking.
The Nb\textsuperscript{+2} state was at its maximum (29\%) when the baking temperature was reached and it slightly dropped (22\%) during baking (Figure \ref{fig:Components_area}(c)).
The Nb\textsuperscript{+1} percentage raised at the time when Nb\textsuperscript{+5} vanished.
Noteworthy that the Nb\textsuperscript{+1} chemical state was not pronounced at \SIrange{200}{230}{\degreeCelsius}.
Contribution of Nb\textsuperscript{+0.4} increased insignificantly and stayed at $\approx$20\%.

At \SI{400}{\degreeCelsius}, the Nb\textsuperscript{+5}–Nb\textsuperscript{+1} states were reduced during the heating stage.
The Nb\textsuperscript{+0.4} state behaved similar to \SIrange{200}{300}{\degreeCelsius}, and it raised as the baking temperature increased (200 °C: 14\%, 230 °C: 16\%, 300 °C-400 °C: 18-23 °C).
The Nb\textsuperscript{+0.6} state appeared simultaneously with Nb\textsuperscript{+1}, reached 18\% and, unlike Nb\textsuperscript{+1}, did not disappear during baking (Figure \ref{fig:Components_area}(d)).
However, the Nb\textsuperscript{+1} state appears upon cooling to room temperature.

Comparing the Nb-O states upon \SI{300}{\degreeCelsius} and \SI{400}{\degreeCelsius} anneals, in the first case the Nb\textsuperscript{+2} state (corresponding to NbO oxide) still remained on the surface while in the latter only the chemical states corresponding to oxygen interstices were left.

The maxima of the relative peak areas in Figure \ref{fig:Components_area} were observed during the heating-up, i.e. when the baking temperature was not reached.
Since similar maxima were observed in the kinetics of the niobium oxide reduction at constant temperatures \autocite{king1990kinetic}, it can be concluded that the maxima are inherent to the kinetics of formation and dissolution processes of niobium oxides.

\subsection{Heat-treated Niobium: Impurities.} 
Analysis of C 1$s$, N 1$s$, P 2$p$ and Si $2p$ core levels showed no evidence of chemical interaction of these elements with niobium at \SI{200}{\degreeCelsius}.

The intensity of Si 2$p$ peak attributed to organic silicon diminished upon baking at \SIrange{200}{300}{\degreeCelsius}, and completely vanished after the \SI{400}{\degreeCelsius} baking (Figure \ref{fig:XPS_Si2p_detailed}(a,b)).

\subsubsection{Phosphorus}
\begin{figure*} 
	\includegraphics[width=\linewidth]{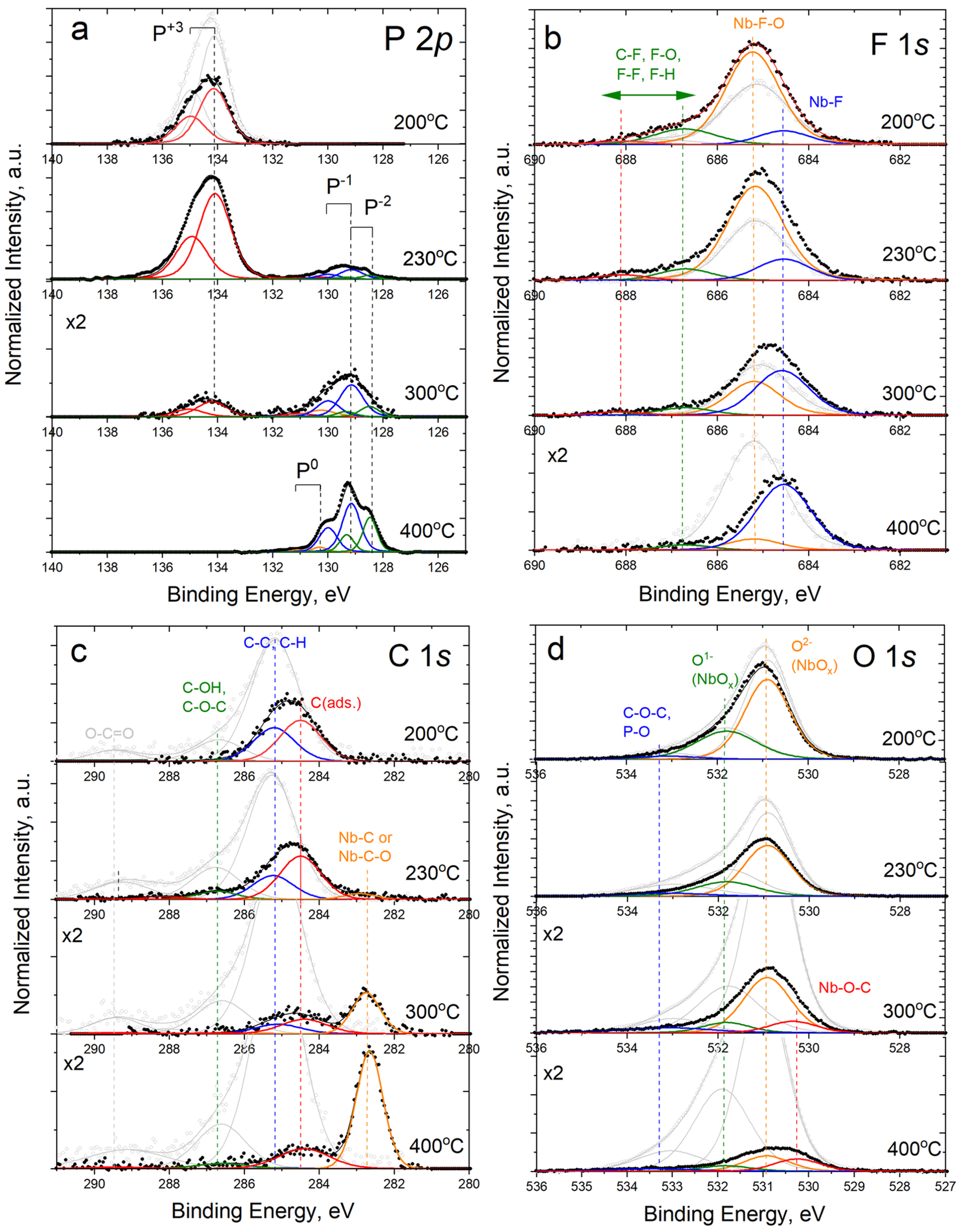}
	\caption{The EDCs of O 1s, C 1s, F 1s ($h\nu$=\SI{1000}{\electronvolt}) and P 2p ($h\nu$=\SI{480}{\electronvolt}) at the end of the respective baking: 200 °C/11.5h, 230°C/15h, 300°C/3h and 400°C/3h. The EDCs are normalized to the area of the Nb 3d core-level.}
	\label{fig:XPS_OCFP_detailed}
\end{figure*}

\hfill\\ 
From the analysis of EDC of P 2$p$ (Figure \ref{fig:XPS_OCFP_detailed}(a)) it follows that the phosphorous-containing species were stable at \SI{200}{\degreeCelsius}, while partially (\SI{230}{\degreeCelsius}, \SI{300}{\degreeCelsius}) or completely (\SI{400}{\degreeCelsius}) reacted with niobium at higher temperatures transforming for the most part to niobium phosphide characterized by P\textsuperscript{-3} (BE=\SI{128.50}{\electronvolt}) and also P\textsuperscript{-2} states (\SI{129.15}{\electronvolt}), as well as free-state phosphorus, P\textsuperscript{0} (\SI{130.25}{\electronvolt}), upon the respective anneal.
Thus, at temperatures above \SI{230}{\degreeCelsius} the chemisorption process of phosphorus was facilitated.
At \SI{200}{\degreeCelsius} the quantity of P\textsuperscript{+3} depleted (probably by desorption) as was determined at $h\nu=\SI{480}{\electronvolt}$ (Figure \ref{fig:XPS_OCFP_detailed}(a)) which was not evident at $h\nu=\SI{1000}{\electronvolt}$ owing to lower surface sensitivity and small concentration of phosphorus. 

\subsubsection{Fluorine}
\hfill\\ 
The position of the main F 1$s$ peak shifts to lower BE by 0, 0.1, 0.2 and \SI{0.6}{\electronvolt} for niobium baked at \SI{200}{\degreeCelsius}/\SI{11.5}{\hour}, \SI{230}{\degreeCelsius}/\SI{15}{\hour}, \SI{300}{\degreeCelsius}/\SI{3}{\hour} and \SI{400}{\degreeCelsius}/\SI{3}{\hour} indicating the change of the chemical state of fluorine during baking.
Additionally, a weak shoulder emerges at $\approx$\SI{687}{\electronvolt}   (Figure \ref{fig:XPS_OCFP_detailed}(b)).
The fitting procedure of F 1$s$ revealed four components positioned at \SI{684.55}{\electronvolt}, \SI{685.15}{\electronvolt}, \SI{686.70}{\electronvolt}, and \SI{688.1}{\electronvolt}. 
In general, metal fluorides reveal peaks in the range of \SIrange{684}{685.5}{\electronvolt} while C-F, O-F, and F-F reveal the peaks at \SIrange{686}{689}{\electronvolt} \autocite{nikolenko2002xps, tressaud1996nature, ninomiya1985reaction}.
Therefore, the  component at \SI{685.15}{\electronvolt} that is present in the initial samples and looses its intensity upon baking, has been interpreted as the Nb-F-O state.
The component at \SI{684.55}{\electronvolt}, on the contrary, increases and is associated with the Nb-F bonds. 
The peaks at \SI{686.70}{\electronvolt} and \SI{688.1}{\electronvolt} can be assigned to the adsorbed fluorine species possibly bonded to oxygen or hydrogen since no peaks associated with fluorine have been observed within C 1$s$.

During baking at \SIrange{230}{400}{\degreeCelsius} new doublets emerged within Nb 3$d_{5/2}$ positioned at around \SI{203.63}{\electronvolt} and \SI{204.30}{\electronvolt} ($\Delta$BE equals \SI{1.63}{\electronvolt} and \SI{2.30}{\electronvolt} respectively).
We suppose that they could be the Nb-O-F and Nb-F states (the estimated quantity of phosphorus is too small to originate sufficient peak-components within Nb 3$d$).

\subsubsection{Carbon}
\hfill\\ 
At \SI{200}{\degreeCelsius}/\SI{11.5}{\hour} the maximum of the C 1$s$  peak was found to be shifted to lower BE upon baking, i.e. a new component at \SI{284.5}{\electronvolt} emerged.
The peak may be referred to the carbon atoms adsorbed to niobium after decomposition of C-C or C-O molecules and be considered as a preceding carbon state prior to the formation of niobium carbide (Figure \ref{fig:XPS_OCFP_detailed}(c)).
Alternatively, the peak at \SI{284.5}{\electronvolt} may be refereed to the carbonaceous species produced by the X-ray beam during measurements. 

After \SI{15}{\hour} at \SI{230}{\degreeCelsius} a small shoulder at $\approx$\SI{282.9}{\electronvolt} could be additionally resolved within C 1$s$ when the spectrum was left for a longer time for accumulation of the signal after the baking (Figure \ref{fig:XPS_OCFP_detailed}(c)).
It could be assigned to the formation of Nb-C or Nb-C-O bonds (because of the small percentage, it was not included in the fitting model for building the plot in Figure \ref{fig:Components_area}(b), but in the analysis of the chemical composition after the baking).
At \SI{300}{\degreeCelsius} and at \SI{400}{\degreeCelsius} the Nb-C peak was located at $\approx$\SI{282.65}{\electronvolt}.
Within the Nb 3$d$ the peak at approximately \SI{203.37}{\electronvolt} was assigned to Nb-C.

It has been established that the interaction of niobium with carbon and other impurities occurs rapidly when the percentage of Nb\textsuperscript{+5} within Nb 3$d$ is below 30\% for 230, 300 and \SI{400}{\degreeCelsius} baked samples, which equals to approximately \SI{1}{\nano\meter} thickness of Nb$_2$O$_5$.

\subsubsection{Oxygen}
\hfill\\ 
It was identified that at \SI{300}{\degreeCelsius} and \SI{400}{\degreeCelsius}, the maximum of the O 1$s$ peak  shifts to lower BE which is associated with the emerging of a new component at \SI{530.3}{\electronvolt} (Figure \ref{fig:XPS_OCFP_detailed}(d)).
It is more intense at \SI{400}{\degreeCelsius} and it gets more pronounced at lower photon energiy of the X-ray beam which implies it originates from the very surface.
This peak may be associated with the formation of Nb-O-C bonds or be a characteristic feature of niobium oxides with lower oxidation numbers.

\subsection{Quantitative Analysis}

The near-surface distribution of elements was estimated in two different ways.

In the first, the data collected at $h\nu$=\SI{900}{\electronvolt} or \SI{1000}{\electronvolt} were used.
In this case the areas of the peaks of the respective EDCs were corrected as described in \autocite{nanotechnology2005ea}.
The advantage of this method is a larger information depth.
Despite the method may have a substantial error it can be used for comparison of the samples. 

In the second approach, the EDCs were acquired at various $h\nu$ providing constant kinetic energy of photoelectrons (\SI{280}{\electronvolt}) and thus equal information depth for all elements \autocite{girard2012chemical}.
The concentration of elements was calculated according to the following expression:

\begin{equation}
	n_i = \frac{S_i}{\sigma_iF_i\displaystyle\sum_{j=1}^{k}\frac{S_j}{\sigma_jF_j}}, 
	\label{eq:concentration}
\end{equation}
where $n_i$ is  the atomic concentration of an element $i$, $S_i$ is the area of EDC originating from a particular core level of the element $i$, $\sigma_i$ is the photoionization cross section of the particular core level of the element $i$, $F_i$ is the photon flux at a particular $h\nu$ at which the core-level of the element $i$ is measured, $k$ is the total number of detected elements.

The first approach is of particular interest for the initial samples, as they have a carbonaceous adsorbate layer on the surface and a thicker niobium pentoxide.
The second method of constant kinetic energy was used additionally to analyze the samples annealed at \SI{300}{\degreeCelsius} and \SI{400}{\degreeCelsius}.
In both approaches the effects related to the angular distribution of the photoelectrons ($L(\beta,\gamma)\approx1$) were minimized by the setup configuration.
The inhomogeneity of the element distribution over depth was not considered. 

An estimate of the near-surface composition of niobium before and after the baking by the first and second approaches is presented in Tables \ref{tab:XPS_concentration_1000ev} and \ref{tab:XPS_concentration_CKE}, respectively. Thus it is possible to write the surface composition ofinitial and baked samples estimated according to the first approach in the following form:
NbO$_{1.99}$F$_{0.20}$N$_{0.04}$ is for the initial niobium and NbO$_{1.64}$F$_{0.3}$ is for the baked at \SI{200}{\degreeCelsius};
NbO$_{1.73}$F$_{0.20}$ → NbO$_{1.03}$C$_{0.02}$F$_{0.35}$P$_{0.01}$ (\SI{230}{\degreeCelsius});
NbO$_{1.91}$F$_{0.16}$N$_{0.04}$ → NbO$_{0.56}$C$_{0.06}$F$_{0.22}$P$_{0.01}$N$_{0.02}$ (\SI{300}{\degreeCelsius});
NbO$_{2.10}$F$_{0.17}$N$_{0.03}$ → NbO$_{0.25}$C$_{0.13}$F$_{0.11}$P$_{0.02}$N$_{0.03}$ (\SI{400}{\degreeCelsius}). 
By the second approach we get the following:
NbO$_{1.38}$F$_{0.08}$N$_{0.01}$ → NbO$_{0.51}$C$_{0.05}$F$_{0.11}$P$_{0.02}$N$_{0.01}$ (\SI{300}{\degreeCelsius});
NbO$_{1.62}$F$_{0.09}$N$_{0.01}$ → NbO$_{0.25}$C$_{0.13}$F$_{0.05}$P$_{0.04}$N$_{0.01}$ (\SI{400}{\degreeCelsius}).
The nitrogen interstitials are also given where they were detectable.

Comparing the initial niobium samples, the estimated amounts of O and F are smaller when they are measured at equal kinetic energy which is primarily related to smaller information depth (\SI{2.3}{\nano\meter} at $h\nu$=\SI{480}{\electronvolt} against \SI{4.6}{\nano\meter} at $h\nu$=\SI{1000}{\electronvolt} in Nb$_2$O$_5$), and as a consequence to presumable probing of surface adsorbates that are always present at the surface prior to heat-treatment.

For the baked samples consistent results were obtained for oxygen and carbon.
However, the constant kinetic-energy approach resulted in a larger amount of phosphorus and a smaller amount of fluorine which may be related to non-homogeneous distribution of these elements over the depth.

\begin{table*}
    \centering
    \caption{Approximate distribution of elements within the near-surface region of the samples, \si{\atpercent} ($h\nu$=\SI{1000}{\electronvolt}).}
    \begin{tabular}{ccccccccc}
        \toprule
        Treatment & Nb & F & O & C & N & Ca & P & Si \\
        \midrule
        \midrule
        \multicolumn{9}{c}{\textbf{\SI{200}{\degreeCelsius}/11.5h anneal}} \\
        Initial & 24.80±0.09 & 5.33±0.29 & 52.24±0.37 & 14.05±0.99 & 1.83±0.49 & 0.33±0.20 & 0.77±0.09 & 0.64±0.21\\
        Nb-bound & 31.00±0.11 & 6.16±0.25 & 61.66±0.34 & 0 & 1.18±0.34 & 0 & 0 & 0\\ 
        \midrule
        Baked & 29.71±0.13 & 10.66±0.33 & 50.48±0.25 & 7.48±0.40 & 0 & 0.28±0.19 & 0.84±0.08 & 0.55±0.18\\
        Nb-bound & 33.99±0.14 & 10.23±0.21 & 55.78±0.22 & 0 & 0 & 0 & 0 & 0\\
        \midrule
        \midrule
        \multicolumn{9}{c}{\textbf{\SI{230}{\degreeCelsius}/15h anneal*}} \\
        Initial & 24.87±0.15 & 5.01±0.17 & 50.24±0.32 & 17.53±0.47 & 0.59±0.20 & 0.40±0.12 & 1.20±0.08 & 0.16±0.16\\
        Nb-bound & 34.14±0.20 & 6.88±0.16 & 58.97±0.39 & 0 & 0 & 0 & 0 & 0\\ 
        \midrule
        Baked & 35.27±0.13 & 14.07±0.19 & 39.05±0.19 & 9.70±0.38 & 0.23±0.12 & 0.34±0.14 & 1.16±0.11 & 0.18±0.16\\
        Nb-bound & 41.60±0.15 & 14.56±0.15 & 42.97±0.18 & 0.62±0.13 & 0 & 0 & 0.26±0.07 & 0\\
        \midrule
        Air & 21.69±0.05 & 1.35±0.13 & 49.86±0.25 & 25.49±0.40 & 0.51±0.15 & 0 & 0.84±0.06 & 0.27±0.12\\    
        Nb-bound & 32.62±0.08 & 1.15±0.10 & 65.66±0.27 & 0.58±0.20 & 0 & 0 & 0 & 0\\  
        \midrule
        \midrule
        \multicolumn{9}{c}{\textbf{\SI{300}{\degreeCelsius}/3h anneal}} \\
        Initial & 26.35±0.23	& 4.33±0.13 & 54.17±0.31 & 12.06±0.64 & 2.20±0.35 & 0 & 0.41±0.09 & 0.47±0.18\\
        Nb-bound & 32.07±0.28 & 5.27±0.16 & 61.41±0.31 & 0 & 1.25±0.24 & 0 & 0 & 0\\ 
        \midrule
        Baked & 48.76±0.35 & 12.04±0.25 & 30.80±0.32 & 6.14±0.44 & 1.56±0.55 & 0 & 0.45±0.14 & 0.25±0.19\\
        Nb-bound & 53.67±0.39 & 11.64±0.17 & 30.15±0.22 & 2.99±0.22 & 1.04±0.40 & 0 & 0.50±0.15 & 0\\
        \midrule
        Air & 25.08±0.08 & 0.97±0.19 & 51.59±0.27 & 19.96±0.48 & 1.56±0.31 & 0 & 0.40±0.08 & 0.45±0.15\\    
        Nb-bound & 32.86±0.10 & 0.92±0.12 & 64.58±0.33 & 0.64±0.20 & 1.00±0.22 & 0 & 0 & 0\\  
        \midrule
        \midrule
        \multicolumn{9}{c}{\textbf{\SI{400}{\degreeCelsius}/3h anneal}} \\
        Initial & 23.93±0.09 & 4.11±0.13 & 53.98±0.29 & 13.68±0.70 & 1.81±0.51 & 0 & 2.26±0.11 & 0.23±0.14\\
        Nb-bound & 30.29±0.11 & 5.21±0.16 & 63.58±0.27 & 0 & 0.92±0.35 & 0 & 0 & 0\\ 
        \midrule
        Baked & 60.21±0.17 & 6.88±0.22 & 17.49±0.63 & 10.96±0.70 & 3.10±0.38 & 0 & 1.36±0.13 & 0\\
        Nb-bound & 65.31±0.18 & 6.96±0.15 & 16.00±0.21 & 8.28±0.23 & 1.97±0.26 & 0 & 1.47±0.14 & 0\\
        \midrule
        Air & 23.50±0.06 & 0 & 44.66±0.22 & 28.25±0.35 & 1.51±0.23 & 0 & 1.69±0.10 & 0.39±0.12\\    
        Nb-bound & 34.25±0.08 & 0 & 61.10±0.25 & 3.33±0.18 & 0.77±0.18 & 0 & 0.54±0.09 & 0\\  
        \bottomrule
        \multicolumn{9}{l}{*\footnotesize{The XPS data were measured at $h\nu= \SI{900}{\electronvolt}$.}}
    \end{tabular}    
    \label{tab:XPS_concentration_1000ev}
\end{table*}

\begin{table*}
    \centering
    \caption{Approximate distribution of elements within the near-surface region of niobium samples, \si{\atpercent}. The XPS spectra were taken at equal kinetic energy of photoelectrons, $E_{kin}$.}
    \begin{tabular}{ccccccccc}
        \toprule
        Treatment & Nb & F & O & C & N & Ca & P & Si \\
        \midrule
        \midrule        
        \multicolumn{9}{c}{\textbf{\SI{300}{\degreeCelsius} anneal}} \\
        Initial & 32.53±0.03 & 2.83±0.11 & 47.71±0.12 & 13.58±0.25 & 0.72±0.11 & 0 & 1.15±0.02 & 1.49±0.03\\
        Nb-bound & 40.48±0.04 & 3.21±0.13 & 56.00±0.14 & 0 & 0.30±0.08 & 0 & 0 & 0\\ 
        \midrule
        Baked & 50.06±0.08 & 6.11±0.18 & 28.51±0.22 & 13.66±0.89 & 0.44±0.16 & 0 & 0.91±0.07 & 0.31±0.07\\
        Nb-bound & 59.14±0.09 & 6.55±0.12 & 29.92±0.16 & 3.06±0.35 & 0.27±0.12 & 0 & 1.08±0.05 & 0\\
        \midrule
        \midrule
        \multicolumn{9}{c}{\textbf{\SI{400}{\degreeCelsius} anneal}} \\
        Initial & 27.88±0.03 & 2.63±0.10 & 46.61±0.15 & 16.82±0.30 & 0.88±0.15 & 0 & 4.04±0.03 & 1.14±0.04\\
        Nb-bound & 36.83±0.04 & 3.47±0.14 & 59.49±0.17 & 0 & 0.22±0.11 & 0 & 0 & 0\\ 
        \midrule
        Baked & 60.47±0.20 & 3.24±0.15 & 17.57±0.31 & 15.86±0.31 & 0.65±0.17 & 0 & 2.21±0.04 & 0\\
        Nb-bound & 68.06±0.23 & 3.47±0.53 & 16.88±0.14 & 8.70±0.10 & 0.40±0.10 & 0 & 2.49±0.04 & 0\\
        \bottomrule        
    \end{tabular}    
    \label{tab:XPS_concentration_CKE}
\end{table*}

It is known that fluorine ions readily enter niobium when it is immersed in HF-acid solution \autocite{antoine1999morphological} and may be located not only in the oxide but also underneath it \autocite{bose2015study}.
According to the presented data, the concentration of fluorine noticeably increased upon baking at \SIrange{200}{300}{\degreeCelsius}, and decreased at \SI{400}{\degreeCelsius}.
The depletion of fluorine could proceed either via evaporation of NbF$_5$ (at \SI{233.5}{\degreeCelsius}) or disproportionation of NbF$_4$(solid) to NbF$_3$(solid) and NbF$_5$(gas) (at temperatures \SIrange{250}{350}{\degreeCelsius}) as described in \autocite{capelli2018thermodynamic}.

The total (i.e. not only bound to Nb) amount of phosphorus was constant at \SI{200}{\degreeCelsius}, while it decreased at \SIrange{230}{400}{\degreeCelsius} likely by desorption as PO$_x$.
The interaction of phosphorus with niobium was noted at \SI{230}{\degreeCelsius}, and got more pronounced at \SIrange{300}{400}{\degreeCelsius}.
The phosphorus species transformed to niobium phosphides completely at \SI{400}{\degreeCelsius}. 
The amounts of carbon and phosphorus that reacted with niobium at \SI{400}{\degreeCelsius} is about twice the amount in the \SI{300}{\degreeCelsius}-sample.

The O/Nb ratio is not equal in the unbaked samples since the initial thickness of Nb$_2$O$_5$ is slightly varied. 
The calculated thicknesses of the top Nb\textsuperscript{+5} oxide only (excluding the oxides having other oxidation states) using the method described in our previous work \autocite{prudnikava2022systematic} for 
the samples before and after the annealing
are presented in Table \ref{tab:Pentoxide_thickness}.
For the \SI{400}{\degreeCelsius}-treated sample the thickness of the remaining Nb\textsuperscript{+2} (NbO) is pointed out.
Thus, upon baking, the thickness of the pentoxide decreases as the temperature of baking increases.

\begin{table} 
	\centering
    \caption{The calculated thickness (in \si{\nano\meter}) of the top niobium oxide (Nb\textsuperscript{+5}) after the respective treatment. The EDC of Nb 3$d$ was acquired at $h\nu$= \SI{1000}{\electronvolt}, except for the \SI{230}{\degreeCelsius}-sample ($h\nu$= \SI{900}{\electronvolt}).}
    \small 
	\begin{tabular}{c c c c}
		\toprule
		  Treatment & Initial & Baked & Air exposed \\
		\midrule
		200°C/11.5h & 2.90±0.07 & 1.57±0.05 & -\\
        230°C/15h & 2.69±0.11 & 0.49±0.01 & 3.36±0.05\\
        300°C/3h & 2.77±0.16 & 0.17±0.04 & 3.15±0.06\\
        400°C/3h & 3.52±0.10 & 0* & 2.32±0.03\\
		\bottomrule
        \multicolumn{4}{l}{*\footnotesize{The thickness of Nb\textsuperscript{+2} is \SI{0.11(1)}{\nano\meter}.}}
	\end{tabular} 	
	\label{tab:Pentoxide_thickness}
\end{table}

\subsection{Air Oxidation}
\begin{figure*} 
	\includegraphics[width=\linewidth]{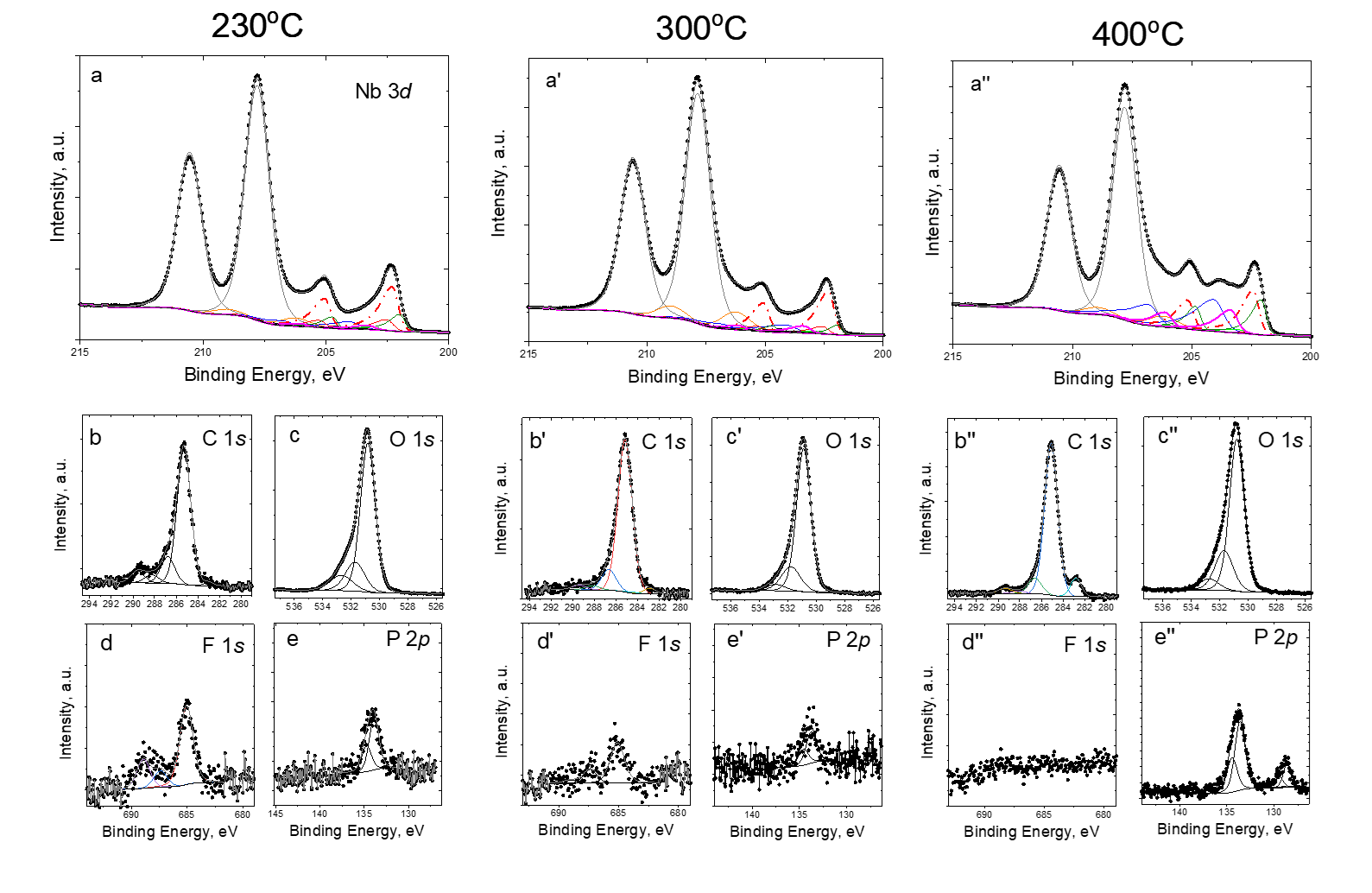}
	\caption{The fitted EDCs ($h\nu=\SI{1000}{\electronvolt}$) of the Nb 3d, C 1s, O 1s, F 1s and P 2p regions of the air-exposed niobium formerly baked at 230 °C/15 h (a–e), 300 °C/3 h (a ’– e’), and 400 °C/3 h (a’’ – e’’). These are non-normalized intensity EDCs fitted with a Shirley background.}
	\label{fig:XPS_after_air}
\end{figure*}

Since the niobium cavities are subjected to air after "furnace baking", we have undertaken a detailed analysis of the air-oxidized niobium after the UHV baking.
After the air exposure for 10 weeks, the XPS spectra ($h\nu=\SI{1000}{\electronvolt}$) of the niobium samples baked at \SI{230}{\degreeCelsius}, \SI{300}{\degreeCelsius} and \SI{400}{\degreeCelsius} were acquired (Figure \ref{fig:XPS_after_air}).

The EDCs of Nb 3$d$  (Figure \ref{fig:XPS_after_air}(a – a’’)) are characterized by an intense Nb\textsuperscript{+5} doublet and thus look similar to the EDCs of the initial BCPed niobium surface.
Analysis of the fitted components within the Nb 3$d$ core-level of the air-exposed annelaed niobium revealed that the binding energy of Nb\textsuperscript{+5} is lower by $\approx$\SI{0.14}{\electronvolt} as compared to the initial niobium (Table \ref{tab:XPS_all_components_air}).
The FWHMs of the oxides did not change so the number of defects did not increase noticeably.  

\begin{table*}
    \centering
    \caption{The binding energy shifts, $\Delta$BE (with respect to Nb\textsuperscript{0}), and the  FWHMs of the fitted components of the Nb 3d core-level obtained for the annealed niobium after the air exposure.}
    \scriptsize
    \begin{tabular}{c c c c c c c c c c c}
        \toprule
        Component & Nb\textsuperscript{+5} & Nb\textsuperscript{+4} &  Nb\textsuperscript{+2} & Nb\textsuperscript{+1} & Nb\textsuperscript{+0.6} & Nb\textsuperscript{+0.4} & Nb\textsuperscript{0} & Nb-C & Nb-O-F & Nb-O-F (II) \\
        \midrule
        BE shift, eV & 5.81±0.03 & 4.18±0.11 & 2.11±0.05 & 1.07±0.07 & 0.61±0.10 & 0.33±0.05 & 0±0.057 & 1.37±0.03 & 1.65±0.05 & 2.39±0.01\\
        FWHM, eV & 1.20±0.05 & 1.42±0.02 & 1.11±0.06 & 0.78±0.04 & 0.67±0.05 & 0.57±0.03 & 0.47±0.04 & 0.63±0.04 & 0.66±0.08 & 0.62±0.08\\ 
        
        \bottomrule
    \end{tabular}    
    \label{tab:XPS_all_components_air}
\end{table*}

The relative areas of the fitted components within the Nb 3$d$ core-level are presented in Table \ref{tab:XPS_all_components}.
The smallest percentage of Nb\textsuperscript{+5} was observed in the \SI{400}{\degreeCelsius}-sample.
The relative area of Nb\textsuperscript{+4} varied between the samples but tended to be larger as compared to the initial niobium.
The minimal relative percentage area of Nb\textsuperscript{+4} measured was 1\%, and its contribution increased during the prolonged XPS measurements as for the initial samples.
Noteworthy, this was observed also for the \SI{400}{\degreeCelsius} sample, where no peak within F 1$s$ was detected after the air exposure.
On the other hand, it has been established that fluorine was not completely eliminated during baking (see Table \ref{tab:XPS_concentration_CKE}).
Based on this, the role of fluorine in the reduction of Nb\textsuperscript{+5} to Nb\textsuperscript{+4} by the X-ray beam can not be completely ruled out.

Comparing to the initial niobium, the Nb\textsuperscript{+2} percentage decreased for the  \SIrange{230}{300}{\degreeCelsius} samples (to \SIrange{3.5}{3.9}{\percent}) while it significantly increased for the \SI{400}{\degreeCelsius} sample (to \SI{12.3}{\percent}).
The contribution of Nb\textsuperscript{+3} and Nb\textsuperscript{+1} states was negligible.
As for the initial niobium, all the samples were characterized by a higher contribution of Nb\textsuperscript{+0.4} as compared to Nb\textsuperscript{+0.6}.

The features inherent to Nb-C state within the EDCs of Nb 3$d$ and C 1$s$ diminished significantly upon air exposure, and were better defined for the \SI{400}{\degreeCelsius} sample due to a higher percentage of Nb-C in the pre-oxidized state (Figure \ref{fig:XPS_after_air}(b – b’’)).

The EDCs of F 1$s$ and P 2$p$ for the air exposed niobium were studied only at $h\nu$=\SIrange{900}{1000}{\electronvolt} (Figure \ref{fig:XPS_after_air}(d-d’’), (e-e’’)).
The intensity of Nb-F peak within F 1$s$ significantly diminished for the \SIrange{230}{300}{\degreeCelsius} treated samples, and the components at a higher BE side interpreted as O-F or H-F bonds were still resolved \autocite{nikolenko2002xps}.
As it has been noted above, no fluorine was detected in the \SI{400}{\degreeCelsius} baked niobium.

The quantitative information on the elemental distribution calculated by the first approach is presented in Table \ref{tab:XPS_concentration_1000ev}.
Owing to presence of the adsorbate layer, the second approach was not used.
The calculated compound formulae are NbO$_{2.01}$C$_{0.02}$F$_{0.04}$, NbO$_{1.97}$C$_{0.02}$F$_{0.03}$N$_{0.03}$ and NbO$_{1.78}$C$_{0.10}$P$_{0.02}$N$_{0.02}$, for \SI{230}{\degreeCelsius}, \SI{300}{\degreeCelsius} and \SI{400}{\degreeCelsius}, respectively.

\subsubsection{Fluorine}
\hfill\\ 
For the niobium baked at \SI{230}{\degreeCelsius} and \SI{300}{\degreeCelsius}, the amount of detected fluorine at the surface after the air oxidation significantly diminished as compared to the just-baked niobium surface ($\approx$ 10 times),
and for the \SI{400}{\degreeCelsius} sample it was not detectable in our experimental conditions (Figure \ref{fig:XPS_after_air}d’’).
The possible reason for the depletion of fluorine in the oxidized samples is the high enthalpy of formation of niobium pentoxide (so fluorine remains in niobium under the oxide layer) and the instability of the formed niobium fluorides in an air environment. 

\subsubsection{Carbon}
\hfill\\ 
After exposure to air, the Nb-C/Nb ratio remained the same in the \SI{230}{\degreeCelsius} sample, while it decreased about 10 times for the \SI{300}{\degreeCelsius} sample, and only 1.3 times for the \SI{400}{\degreeCelsius} sample (Table \ref{tab:XPS_concentration_1000ev}).
Thus, the amount of  the reacted carbon within the XPS information depth in the \SI{400}{\degreeCelsius} sample after the air exposure exceeded the analogous for the \SI{300}{\degreeCelsius} sample by 20 times, while right after the baking they differed only 2.3 times.

One could suggest that at \SI{300}{\degreeCelsius} a portion of the detected Nb-C bonds was also shared with oxygen as Nb-C-O, and upon air oxidation a portion of these oxygen atoms participated in the formation of niobium oxide.

\subsubsection{Phosphorus}
\hfill\\ 
In the \SI{230}{\degreeCelsius} and \SI{300}{\degreeCelsius} baked samples both the free (P\textsuperscript{0}) and the bound phosphorus (P\textsuperscript{-2}, P\textsuperscript{-3}) that were formed during the baking transformed to the P\textsuperscript{+4} state (BE(P $2p_{3/2}$)=\SI{134.4}{\electronvolt})  upon air exposure.
Thus, no niobium phosphides were detected in these samples.
In the \SI{400}{\degreeCelsius} sample, the P\textsuperscript{-3} state still remained along with the P\textsuperscript{+4} state.
The total P/Nb ratio slightly decreased as compared to the unbaked samples: from 0.05 to 0.04 and from 0.09 to 0.07 for the \SI{230}{\degreeCelsius} and \SI{400}{\degreeCelsius} samples, respectively. 
It was not obvious for \SI{300}{\degreeCelsius} (P/Nb=0.02 before and after the air exposure), since the quantity of this element was relatively small in the initial sample.
Also, a small non-uniformity in the distribution of P over the surface could be present. 
It was checked whether the P-containing particle agglomerates could influence the measured photoelectron current between different measurement runs.
For this purpose several areas have been tested by XPS.
The SEM/EDX inspection also has not revealed such particles.
For example, for the \SI{400}{\degreeCelsius} sample the EDX mapping in several locations revealed the P/Nb-ratio of 0.12–0.02 at accelerating voltages of the electron beam in the range \SIrange{3}{10}{\kilo\volt}, correspondingly.

\subsubsection{Oxygen}
\hfill\\ 
The amount of oxygen in the samples is determined substantially by the formed surface oxide layer.
The calculated thicknesses of the top Nb$_2$O$_5$ layer using the data collected at $h\nu=\SI{1000}{\electronvolt}$ for \SI{230}{\degreeCelsius}, \SI{300}{\degreeCelsius} and \SI{400}{\degreeCelsius}-baked niobium followed by air exposure are presented in Table \ref{tab:Pentoxide_thickness}.
These values correlate with the amount of carbon chemically bound to niobium, i.e. the more Nb-C percentage, the smaller Nb\textsuperscript{+5} and larger Nb\textsuperscript{+2} contributions.
Fluorine effect was hard to establish.

\subsection{Kinetics of Oxide Reduction }
\begin{figure} 
	\includegraphics[width=\linewidth]{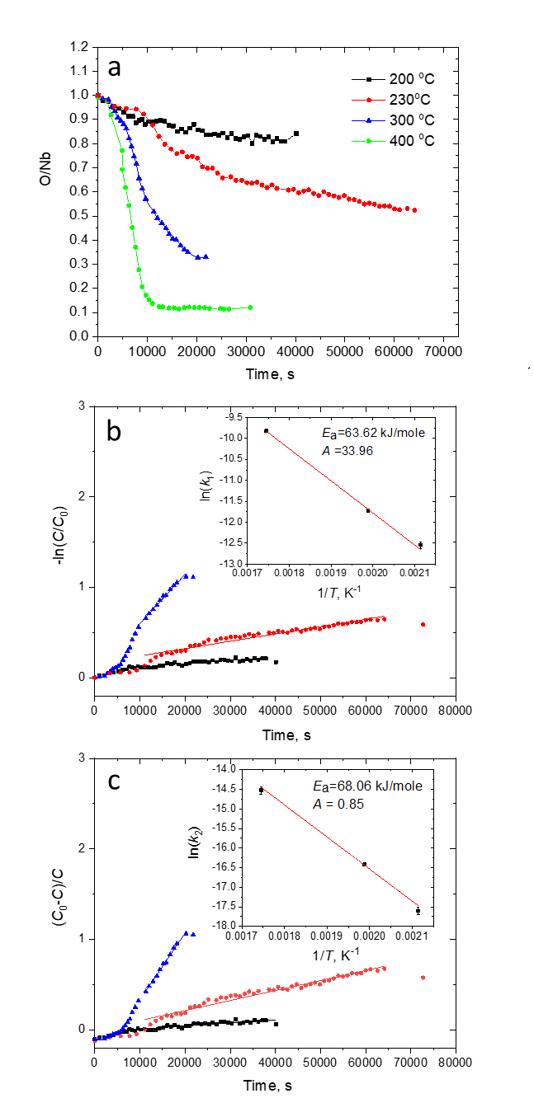}
	\caption{Kinetics of the oxide-layer reduction process: (a) the time dependence of the normalized ratio of O 1s and Nb 3d peak area at various temperatures. (b, c) Fits to the kinetic models for the first- and second-order reactions, respectively. Arrhenius plots created using the calculated rate constants are shown in the insets.}
	\label{fig:O_to_Nb}
\end{figure}

The process of niobium oxide reduction via oxygen dissolution into niobium was studied by the analysis of the ratio of peak areas corresponding to oxygen which is bound to niobium (fitted within O 1$s$) and the total area of niobium (fitted within Nb 3$d$) as a function of annealing time (Figure \ref{fig:O_to_Nb}(a)). 

For the analysis, only the XPS data collected at the temperature of baking were accounted (the data gathered during the heating-up stage were not used).
Since the oxide layer was completely decomposed during the heating-up before reaching \SI{400}{\degreeCelsius}, this temperature was not considered.
In the studied temperature range of \SIrange{200}{300}{\degreeCelsius}, the data fit both the kinetics of the first- and second-order reactions. 
The corresponding data plotted in coordinates $ln\frac{C}{C_O}$ and $\frac{C_O-C}{C}$ against time $t$ are shown in Figures \ref{fig:O_to_Nb}(b) and (c), respectively.
Probably, since the oxide decomposition started in some cases earlier than the target temperature was reached, the deviation from these models is observed in the beginning.

The dependencies $lnk(T^{-1})$ for the baking experiments at temperatures \SIrange{200}{300}{\degreeCelsius} are shown in the insets of Figures \ref{fig:O_to_Nb}(b, c).
It follows that the rate constants for the first- and second-order reactions are determined as $k_1=33.96exp({\frac{-63.62\pm2.49}{RT}})$ and $k_2=0.85exp({\frac{-68.07\pm4.97}{RT}})$.
These relations represent the Arrhenius equations with the activation energy, $E_a$, in \si{\kilo\joule\per\mol}. 
The obtained values of $E_a$ (\SI{63.6}{\kilo\joule\per\mol} and \SI{68.1}{\kilo\joule\per\mol}) are much smaller as compared to \SI{174}{\kilo\joule\per\mol}  obtained in \autocite{king1990kinetic} and \SI{119.9}{\kilo\joule\per\mol} \autocite{lechner2021rf}, but similar to \SI{58.2}{\kilo\joule\per\mol} in \autocite{veit2019oxygen}.
For comparison, in Figure \ref{fig:O_rate_constants} the temperature dependence of the rate constants for the niobium oxide dissolution reaction of the first, $k_1$, and second, $k_2$,  orders obtained in the present work are plotted together with the published values.

\subsection{The Oxygen Concentration Depth Profile}
Knowing the rate constants for oxide reduction it is possible to calculate the oxygen-concentration depth profiles in niobium.

\subsubsection{200-300°C}
\hfill\\ 
In the cases of annealing at temperatures \SIrange{200}{300}{\degreeCelsius} and the durations applied, the following solution of the second Fick’s law can be applied (“the continuous plane source”) \autocite{jaeger1959conduction} (p.262, eq.8):

\begin{equation}
	C = \frac{1}{\sqrt{\pi D}}\int_{0}^{t}\frac{e^{-x^2}}{4Dt}\frac{\phi(t)dt}{\sqrt{t}}, 
	\label{eq:plane_source}
\end{equation}

where $D$ is the diffusion coefficient, \si{\centi\meter\squared\per\second}, $t$ is the duration of baking, $\phi(t)$ is the rate at which the diffusing species are liberated from the surface plane, i.e. the rate of oxide reduction. 
As compared to \autocite{jaeger1959conduction}, “2” is excluded from the denominator as Equation \ref{eq:plane_source} is applied for a semi-infinite body.
At the duration of baking used in our experiments $\phi(t)=v$, i.e. is constant, so Equation \ref{eq:plane_source} becomes:

\begin{equation}
	C = v\sqrt{\frac{t}{\pi D}} exp({\frac{-x^2}{4Dt}}) - \frac{vx}{2D} erfc(\frac{x}{2\sqrt{Dt}}) , 
	\label{eq:plane_source_concentration}
\end{equation}

\begin{figure} 
	\includegraphics[width=\linewidth]{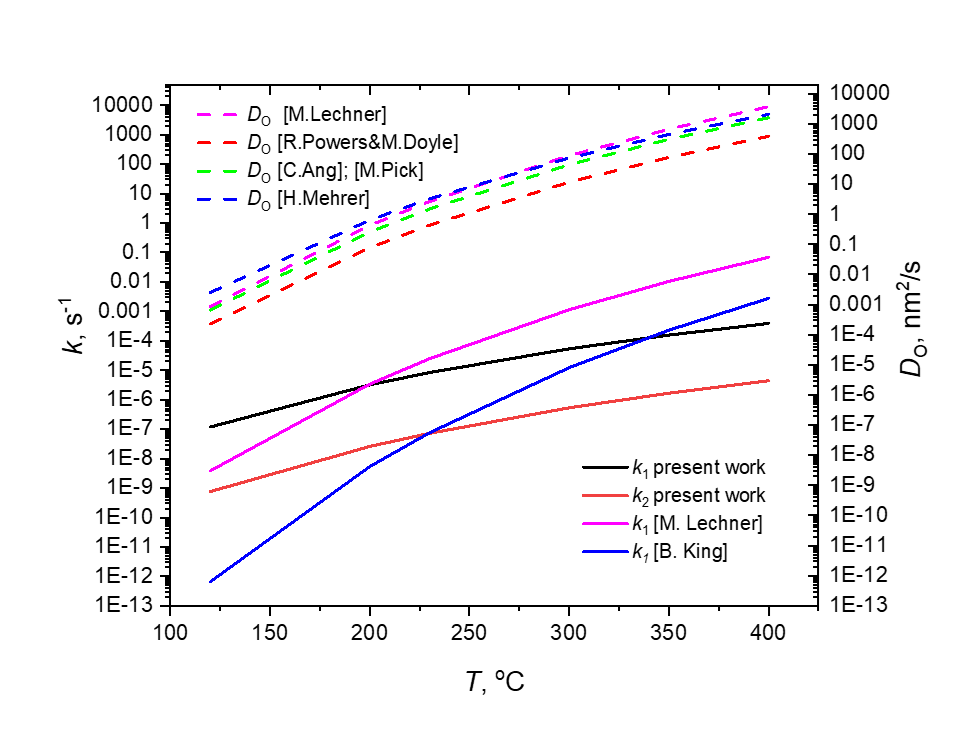}
	\caption{Temperature dependence of the rate constants for the niobium oxide reduction of the first, $k_1$, and the second order, $k_2$, obtained in the present work in comparison with the published values \autocite{ lechner2021rf, king1990kinetic}, as well as the oxygen diffusion coefficients, $D_0$, from different references \autocite{lechner2021rf, pick1982depth, powers1959diffusion, ang1953activation}.}
	\label{fig:O_rate_constants}
\end{figure}

\begin{figure} 
	\includegraphics[width=\linewidth]{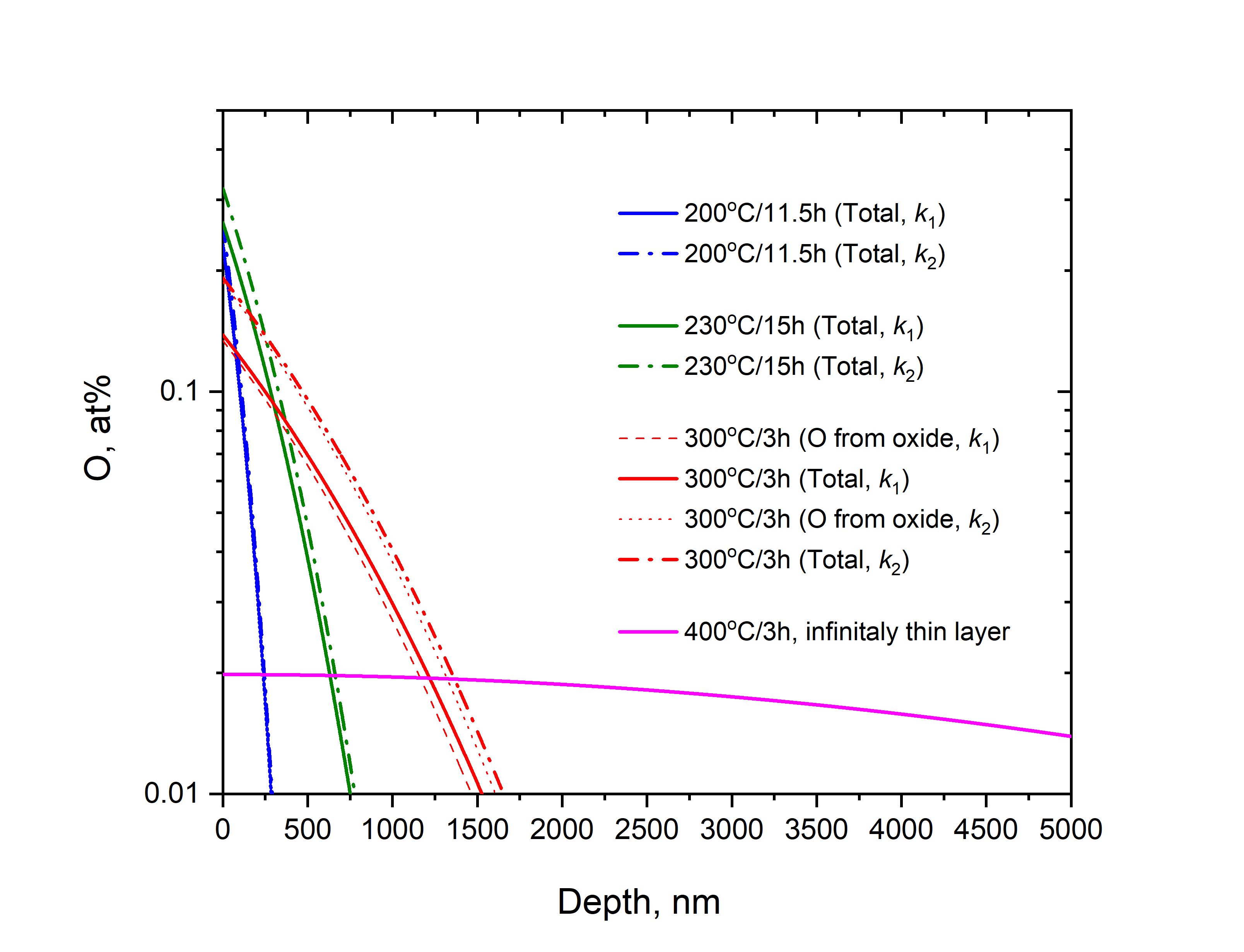}
	\caption{The calculated concentration profiles of oxygen in niobium upon respective anneal using the obtained $k_1$ and $k_2$ and $D_0$ from \autocite{pick1982depth, ang1953activation}. Dashed lines: oxygen was liberated from the oxides (Nb$_2$O$_5$ / NbO$_2$ / Nb$_2$O$_3$ / NbO) only, $k_1$(dash-dot: $k_2$) was used. Solid lines: the O-interstitial concentration enrichment (average value of 5.3 at\%) under the oxides in the initial Nb is additionally considered, $k_1$(dot-dot: $k_2$) was used.
 }
	\label{fig:O_profiles}
\end{figure}

The oxygen diffusion coefficient was taken from \autocite{pick1982depth} and is shown in Figure \ref{fig:O_rate_constants} for comparison to other referenced values.
The results of calculation of the depth profiles of oxygen under the surface oxides using both $k_1$ and $k_2$ are presented in Figure \ref{fig:O_profiles}.
The curves were calculated using $v_1 = k_1C_0$ and $v_2 = k_2C_0^2$, where $C_0$ is the average concentration of oxygen in the oxide layer among the studied initial Nb samples.
The contribution from a \SI{5}{\atpercent}-oxygen enriched layer below the oxide layer was also accounted and was negligible.
Oxygen profile caused by the oxide only (without contribution of the oxygen-enriched layer) is shown in Figure \ref{fig:O_profiles} with a dashed line for the \SI{300}{\degreeCelsius} sample.
The calculated amount of liberated oxygen from the oxide agrees with the XPS data (Table \ref{tab:Oxygen_components}).
With $k_1$ a better agreement was obtained. 
The obtained curves represent the depth profiles of oxygen under the remaining surface oxides.

\subsubsection{400°C}
\begin{figure} 
	\includegraphics[width=\linewidth]{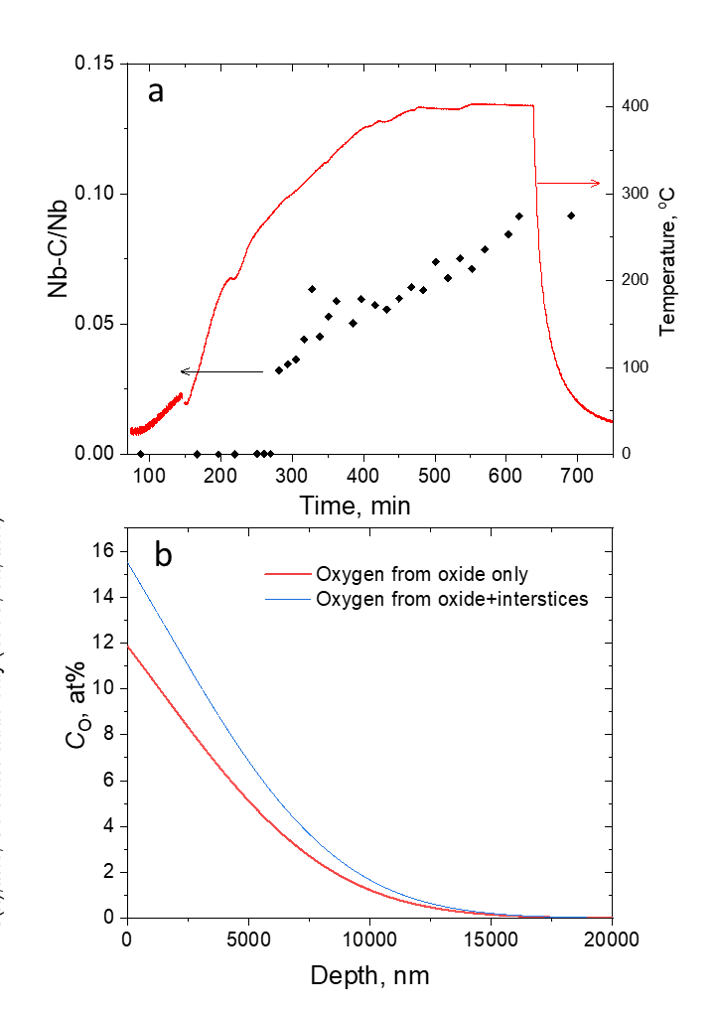}
	\caption{(a) The Nb-C/Nb ratio estimated by fitting the peak area within C 1s and Nb 3d ($h\nu$=\SI{900}{\electronvolt}) during the thermal anneal at \SI{400}{\degreeCelsius}. (b) The hypothetical oxygen concentration profile at \SI{400}{\degreeCelsius} if the  partial oxygen pressure was \SIrange{6e-4}{8e-4}{mbar}.}
	\label{fig:concentration_400C}
\end{figure}

\hfill\\ 
The oxide-dissolution process was completed during the heating-up before \SI{400}{\degreeCelsius} was reached.
No gain in the O/Nb ratio was observed during \SI{3}{\hour} baking time (O/Nb = 0.25).
A slight increase in the Nb-C/Nb ratio was observed both during the heating-up stage and constant temperature baking (Figure \ref{fig:concentration_400C}(a)).

At \SI{400}{\degreeCelsius} an oxygen-solute atom may travel within the Nb lattice at a rate of \SI{40}{\nano\meter\per\second}, which is about 6 times faster than at \SI{300}{\degreeCelsius}.
One could assume that the surface-concentration level of oxygen is kept constant since its diffusion into the bulk is compensated by absorbing atoms from the ambient environment.
However, considering the UHV condition this option can hardly be the case.
It is possible to roughly estimate the maximal gain in solute concentration of gas atoms, $\Delta C$, which depends on the number of gas molecules striking the surface per unit time.
According to \autocite{fromm1980hydrogen}:

\begin{equation}
	\Delta C \approx \frac{1}{2} \frac{pt}{d} , 
	\label{eq:concentration_gain}
\end{equation}
where $\Delta C$ is the change of the concentration of dissolved species in \si{\atpercent}, $p$ is the residual pressure of reactive gases in \si{\milli\bar}, $t$ is the heat treatment duration in \si{\second}, and $d$ is the thickness of the sample in \si{\centi\meter}.
Thus, at a pressure of \SI{3.5e-9}{\milli\bar} of oxygen-containing species in the vacuum chamber at \SI{400}{\degreeCelsius}, the maximal $\Delta C$ after \SI{3}{\hour} is estimated to be \SI{6.75e-5}{\atpercent}.
The real $\Delta C$ is usually smaller since the formulae does not account for surface reactions.
The obtained value of $\Delta C$ is an order of magnitude smaller than the bulk oxygen concentration of niobium with $RRR$=300 which is below \SI{6.78e-4}{\atpercent} (i.e. below 10 ppm by weight according to the material specifications) and thus is negligible in our case.

However, if the base pressure was higher, $\Delta C$ would rise and the constant surface concentration mode described by $C=C_s erfc(\frac{x}{2\sqrt{Dt}})$ (where $C_s$ is surface concentration) could be realized during baking.
This situation would have taken place at a base oxygen pressure of \SIrange{6e-4}{8e-4}{mbar} 
considering the estimated surface concentration of $C_0$ = 15.49 (11.86) at\% (if the oxygen sustained both in oxides and interstitials (only interstitials) is counted).
This hypothetical case is demonstrated in Figure \ref{fig:concentration_400C}(b).

Since $\Delta C$ is quite low, the oxide dissolution rate and the oxygen-diffusion coefficient are high, so that one can disregard the kinetics of oxide dissolution and consider  solving the diffusion equation for the infinitely thin layer:

\begin{equation}
	C = M \frac{1}{\sqrt{\pi Dt}}exp(\frac{-x^2}{4Dt}) , 
	\label{eq:thin_layer}
\end{equation}
where $M$ is the number of diffusing species per unit area.
The calculated surface concentration of $\approx$\SI{0.02}{\atpercent} is much lower than the measured by XPS (Figure \ref{fig:O_profiles}).
This is explained by the presence of a thin oxygen-segregation layer with oxide precipitates (in our case, trace amounts of NbO and the interstices), typical for exothermal metal-gas system such as Nb-O at the temperatures under investigation \autocite{horz1983role}, which is being probed by XPS.

\section{Summary and Discussion}
\label{summary}
To summarize, the initial BCPed niobium surface was composed of Nb, O, F with surface impurities containing P, O and C in the form of phosphate species, most likely (H$_2$PO$_4$)\textsuperscript{-} ions, and carbonaceous contamination.
The identification of the chemical states of niobium was performed by analysis of the Nb 3$d$ core-level using a well-established fitting model \autocite{prudnikava2022systematic} based on the known chemical shifts of niobium in various oxides.
Expectedly, niobium surface was dominated by Nb\textsuperscript{+5} state which corresponds to Nb$_2$O$_5$ oxide, the most stable in normal conditions \autocite{earnshaw1997chemistry}.
As it has been mentioned, the oxygen variation between the samples is governed by the variation of the native oxide thickness.

Previously it was reported that the composition of the native oxide grown in either air or oxygen at atmospheric pressure at the clean Nb surface \autocite{halbritter1987angle, ma2003angle, delheusy2008x}, or after the BCP \autocite{ma2004thermal} is represented by a dominant Nb\textsuperscript{+5} (Nb$_2$O$_5$) as well as Nb\textsuperscript{+4} (NbO$_2$), Nb\textsuperscript{+2} (NbO), and metal-rich sub-oxides.
Our measurements reveal that the native oxide grown at the surface of the BCPed niobium and stored in air environment for several days is composed of Nb\textsuperscript{+5} (Nb$_2$O$_5$), Nb\textsuperscript{+2} (NbO) oxides and oxides of lower oxidation states.
It has been established that the initially absent Nb\textsuperscript{+4} state emerges under the X-ray photon beam.

It has been revealed that the native oxide contains oxyfluoride.
Noteworthy, the fluorine was not detected in the BCPed niobium by XPS in previous works \autocite{ma2003angle, ma2004thermal}.

\subsubsection{Presence of Fluorine}
\hfill\\ 
The sources of fluorine in niobium lattice is the hydrofluoric acid which is a constituent of the BCP solution.
Fluorine ions enter the niobium lattice during etching. 
Niobium and its oxides undergo fluorination even at room temperature due to a strong oxidizing ability of fluorine  \autocite{agulyansky2004chemistry}.
During the BCP, niobium and fluorine may form a number compounds and complexes in HF-aqueous solution: NbO$_2$F, NbF$_5$, [NbOF$_4\cdot$H$_2$O]\textsuperscript{–}, [NbOF$_5$]\textsuperscript{2–}, [NbF$_7$]\textsuperscript{–}, [NbF$_6$]\textsuperscript{–} \autocite{lu2014solution}.
Some of them may subsequently precipitate in different compounds upon water or air exposure. 
The type of transformation and the final compound depend on various factors (humidity, temperature, etc.).
Besides, the niobium surface is being oxidized by oxygen during the chemical treatment and during the rinsing steps.
The oxidation continues when the surface is exposed to air.
Fluorine remains in the lattice during these steps and thus can not be removed by a water rinsing and HPR.

Identification of the  BE shift of fluorine within the Nb 3$d$ is complicated owing to numerous Nb-oxidation states of in its native oxide.
The following aspects may a little elucidate the state of fluorine.
Firstly, the very first collected spectra of Nb 3$d$ measured at $h\nu$=\SIrange{900}{1000}{\electronvolt} did not contain Nb\textsuperscript{+4} component which is usually attributed to NbO$_2$ oxide.
During the XPS measurement at room temperature, a part of the Nb\textsuperscript{+5} state got reduced to Nb\textsuperscript{+4} (gaining up to $\approx$4\% of relative area within Nb 3$d$ which is close to $\approx$\SI{5}{\atpercent} of F content in the BCPed Nb samples) during several hours of illuminating  the \SI{1000}{\electronvolt} X-ray beam.
Thus, the photon beam promoted the chemical changes in the Nb-O-F system.
At that, a shoulder appeared at \SIrange{686.8}{686.7}{\electronvolt} within F 1$s$.
It is to be explored whether the X-ray photon beam alone may provoke the reduction of niobium oxide if it did not contain fluorine in its lattice.

From these data it follows that fluorine is distributed at least within the surface layer equal to the XPS-information depth, and thus is being embedded in the native oxide. 
At room temperature, Nb\textsuperscript{+4} may be interpreted as Nb-F or more likely Nb-O-F-bonding \autocite{wu2019fluorine}. 
The possibility that Nb-O-F bonding exists in oxides of lower oxidation states (from Nb\textsuperscript{+3} to Nb\textsuperscript{+0.4}) is not excluded but at least at room temperature their content is negligible.
Since the new weak peak features appear within F 1$s$, we deem that in our experimental conditions a tiny portion of fluorine ions has overcome the potential-surface barrier and transformed to a surface adsorbate as H-F or C-F molecules \autocite{capelli2018thermodynamic}.
Complementary, fluorine may have been redistributed within the interstitial (octahedral \autocite{agulyansky2004chemistry}) positions of Nb(oxide) lattice.

It was previously reported that a slight decomposition of niobium oxyfluoride, particularly Nb$_3$O$_7$F, occurs readily in the beam of an electron microscope \autocite{bursill1973direct}.
Such transformations upon low-beam heating may proceed via dislocation mechanism involving also grain boundaries.
Similarity of the ionic radii of oxygen and fluorine ensures easy substitution of these atoms with each other in the metal lattice.
It was shown that fluorine-oxygen substitution yielding oxyfluoride compounds can be performed both without changing and with cardinal changes of the initial oxide structure \autocite{agulyansky2004chemistry}.
These facts may explain the embedding of fluorine atoms in the oxide structure.

Since fluorine atoms easily move in the interstitial positions of the niobium-oxide lattice, then during baking the increase of Nb\textsuperscript{+4} may be related to both the process of Nb\textsuperscript{+5} reduction, and the formation of Nb-F states \autocite{capelli2018thermodynamic}.
It is likely, that during baking F\textsuperscript{–} ions are also located beyond the pentoxide layer, and may also overlap with the peaks assigned to niobium oxides.

\subsubsection{Niobium Surface upon Baking Treatments}
\hfill\\ 
A detailed analysis of the surface state of niobium upon annealing at \SI{200}{\degreeCelsius}, \SI{230}{\degreeCelsius}, \SI{300}{\degreeCelsius} and \SI{400}{\degreeCelsius} has been presented.
Briefly, at \SI{200}{\degreeCelsius}, neither P nor C reacted with niobium even after the prolonged anneal.
The amount of adsorbed carbonaceous contamination decreased by approximately 50\% due to desorption process. 
The surface was dominated by the Nb\textsuperscript{+5} chemical state with approximately equal percentages of Nb\textsuperscript{+4}, Nb\textsuperscript{+2}, Nb\textsuperscript{+0.4}, Nb\textsuperscript{0} (Nb\textsuperscript{+3} Nb\textsuperscript{+1} Nb\textsuperscript{+0.6} were subtle).
The thickness of the upper oxide decreased from \SI{2.89}{\nano\meter} to \SI{1.56}{\nano\meter}.

At \SI{230}{\degreeCelsius} some interaction with carbon (Nb-C/Nb=0.015) and phosphorus (Nb-P/Nb=0.006) was detected upon the \SI{15}{\hour} anneal that was cross-checked with a lower $h\nu$ providing higher surface sensitivity.
The Nb-F to Nb ratio increased  $\approx$1.75 times.
The resulted niobium surface at \SI{230}{\degreeCelsius} was represented by Nb\textsuperscript{+5}, Nb\textsuperscript{+4}, Nb\textsuperscript{+3}, Nb\textsuperscript{+2}, Nb\textsuperscript{+0.4}, Nb\textsuperscript{0} oxidation states in an approximately equal proportion.
Thus, a new chemical state (Nb\textsuperscript{+3}) emerged.
The thickness of the layer corresponding to Nb\textsuperscript{+5} changed from \SI{2.68}{\nano\meter} to \SI{0.49}{\nano\meter}.

In the kinetics of Nb 3$d$ components at \SIrange{200}{230}{\degreeCelsius} the dissolution rate of Nb\textsuperscript{+5} changes with time and it is slightly higher at the beginning of the treatment (it was not observed at \SIrange{300}{400}{\degreeCelsius}, at least at the heating rates applied).
There are several possible explanations for this observation.
One can suggest that this can be related to the heating process, which causes the increase of the base pressure in the chamber, thus increasing the sticking probability of gaseous species to the adsorption centers at the sample surface, their subsequent absorption and diffusion, thus promoting the oxide dissolution. 

In \autocite{bose2020evolution} it was suggested that at temperatures above \SI{100}{\degreeCelsius} a transformation of Nb$_2$O$_5$ to NbO$_2$ is favored by reaction of Nb$_2$O$_5$ with a graphitic layer.
This supposes the removal of surface carbon as CO or CO$_2$ by consuming oxygen atoms from Nb$_2$O$_5$.
However, the desorption of oxygen as CO as a product of Nb$_2$O$_5$ reduction to NbO$_2$ by graphitic carbon was previously observed experimentally at \SIrange{1050}{1250}{\kelvin} using solid pellet-type samples \autocite{taylor1967solid}.
So, this scenario is unlikely feasible in our case.
On the other hand, it can be explained by fluorine redistribution in niobium, possible exchange of fluorine atoms with oxygen in an oxide sublattice, which was already observed during the heating and the photoionization process caused by the photon beam.

Finally, the increased dissolution rate of pentoxide may occur due to thermotransport and electrotransport in addition to a diffusion flux, $D\nabla C$.
Thus, the overall flux of oxygen can be described by \autocite{fromm1980hydrogen}:

\begin{equation}
	J = -D\nabla C + \frac{DC}{R}Q^*\nabla\frac{1}{T}+\frac{DC}{RT}FZ^*\nabla\Phi,
	\label{eq:total_oxygen_flux}
\end{equation}
where $\nabla C$, $\nabla\frac{1}{T}$, $\nabla\Phi$ are the concentration gradient, the gradient of reciprocal temperature, and the electric potential gradient, respectively;  $Q$\textsuperscript{*} is the heat of transport, $Z$\textsuperscript{*} is effective valence.

After the \SI{300}{\degreeCelsius} baking the surface was represented by all possible oxidation states of niobium.
It was characterized by Nb\textsuperscript{+0.4}(22\% relative area) and Nb\textsuperscript{+2} (NbO, ~19\%) followed by Nb\textsuperscript{+1} (Nb$_2$O, 14\%) with trace quantity of higher oxides (Nb\textsuperscript{+5}, Nb\textsuperscript{+4} and Nb\textsuperscript{+3}).
The maximal percentage of Nb\textsuperscript{+0.4} was observed for this treatment.
The resulted surface concentration of niobium fluorides significantly exceeded the concentration of niobium carbides (Nb-F/Nb=0.22 vs. Nb-C/Nb=0.056, or \SI{11.6}{\atpercent} vs \SI{3.0}{\atpercent}, respectively) for this sample, and it is only 2.55 times less than the concentration of oxides.
Thus, the role of fluorine has to be considered when the impact of surface impurities onto the superconducting properties of niobium is discussed.
The thickness of the pentoxide was equal to \SI{0.17}{\nano\meter}. 

At \SIrange{200}{300}{\degreeCelsius} (at least at the duration of baking and base pressures explored), along with increase of fluorine concentration, the doping of niobium with oxygen proceeds via oxide dissolution.
In other words, the native oxide is the only source of oxygen atoms (the absorption of the gaseous species from vacuum is negligible).
Some interaction of carbon with niobium occurred at \SIrange{230}{300}{\degreeCelsius} which in the form of carbide or oxycarbide was allocated at the surface since carbon occupies the same interstitial positions (octahedral) in niobium lattice as oxygen \autocite{vishwanadh2016formation} and its diffusion coefficient is at least two orders of magnitude smaller as compared to oxygen.
According to our results, the interaction of carbon with niobium becomes pronounced when the relative area of Nb$_2$O$_5$ within Nb 3$d$ drops below 30\% which equals to the Nb$_2$O$_5$ thickness of $\approx$\SI{1}{\nano\meter}.
Thus, making the top oxide a few nanometers thicker would help to decrease or even eliminate the amount of reacted carbon, and also phosphorus, as well as similar kind of impurities.
This would open possibilities for studying the sole effect of oxygen-doping profile modified by various temperatures and durations on the cavity performance excluding the influence of precipitates of foreign phases.

At \SI{400}{\degreeCelsius} the vacuum-baked niobium surface is dominated by pure niobium Nb\textsuperscript{0} (34\%) and interstitials (Nb\textsuperscript{+1}, Nb\textsuperscript{+0.6} and Nb\textsuperscript{+0.4}).
Only the amount of Nb\textsuperscript{+2} (NbO) (<5\% relative area within Nb 3$d$) equivalent to \SI{0.1}{\nano\meter}-thick layer was left.
The quantity of Nb-F decreased as compared to the initial concentration (Nb-F/Nb=0.11 vs 0.17) and was the lowest among the treated samples (\SI{6.96}{\atpercent}), while carbide was also formed and was maximal for this treatment (Nb-C/Nb=1.13, or \SI{8.28}{\atpercent}).
It is important to emphasize that the top oxide was dissolved before reaching the baking temperature, and afterward, O/Nb ratio did not change with time (Figure \ref{fig:O_to_Nb}(a)).
Since the oxygen diffusion coefficient in niobium at \SI{400}{\degreeCelsius} is pretty high, oxygen can be absorbed by niobium during baking and the gain of oxygen would be determined by its partial pressure in the vacuum chamber and the decomposition rate of oxygen-containing molecules that constitute the residual gas at the niobium surface.
Thus, in this regime, a lot more oxygen than it is retained in the native oxide may
potentially be absorbed by niobium.
In this way, along with NbC$_x$ formation and possible interaction with “native” furnace contamination (specific to a particular chamber), uncontrolled loading of oxygen into niobium, determined by the residual partial pressure in the vacuum furnace, is the most probable reason of poor cavity performance treated with this recipe \autocite{ito2021influence}.
This fact has to be considered when calculating a diffusion profile of impurities using the Fick’s Law and the empirical formulae predicated on the specific boundary conditions. 
In particular, the Equation \ref{eq:plane_source_concentration} used for calculating the oxygen profile in niobium by dissolution of niobium oxide \autocite{ciovati2006improved, lechner2021rf} is valid at temperatures up to \SI{300}{\degreeCelsius}, and must not be used at \SI{400}{\degreeCelsius} as suggested in \autocite{lechner2021rf, khanal2023insight}.

\subsubsection{Other Aspects of Baking Treatments}
\hfill\\ 
Some carbon adsorbates are still present even at the end of \SIrange{300}{400}{\degreeCelsius} baking, and their amount increases during accumulation of the XPS spectra. This is explained by the decomposition of organic species at the sample spot where the photon beam shines.

In some samples trace amounts of Ca-O and organic silicon were found.
Few Ca-O particles were also identified with SEM/EDX.
These impurities neither affected the chemical reduction process of niobium oxides during baking nor the interaction of Nb with C- and P-containing surface contamination.

\subsubsection{Baked Niobium after the Air Exposure}
\hfill\\ 
Upon subjecting the vacuum-baked samples to the air environment, the niobium surface gets oxidized and the amount of fluoride and carbide diminish at the surface.
Niobium carbides in a form of bulk material or powder are more persistent to oxidation as compared to pure niobium, and their oxidation is known to occur at temperatures above \SI{400}{\degreeCelsius} \autocite{shevchenko1980oxidation, shimada1993kinetic}.
Here the oxidation occurs at room temperature, so we assume that part of Nb-C bonds breaks off during reconstruction of niobium lattice as oxides are formed.
As to fluorine, it easily exchanges with oxygen in niobium lattice owing to equal ionic radii, and probably less well with carbon since it has a larger ionic radius as compared to O and F ions.
So, after the air exposure both fluorine and carbon would be allocated below the oxide layer. The fluorine concentration profile upon such bakings is yet to be explored.

The BE of Nb\textsuperscript{+5} of the baked niobium surface oxidized in air was lower by about \SI{0.14}{\electronvolt} as compared to the BCPed niobium.
Our additional studies of the niobium-oxide growth at the Nb surface after the \SI{300}{\degreeCelsius} baking upon air exposure showed that $\Delta$BE of Nb\textsuperscript{+5} with respect to Nb\textsuperscript{0} increases with time (from \SI{5.35}{\electronvolt} for \SI{5}{\hour} of air exposure to \SI{5.45}{\electronvolt} for \SI{12}{\hour}).
One could think that the pentoxide grown in wet condition (water rinsing after the BCP) is “better oxidized” as compared to the one grown in air.
On the other hand, the presence of fluorine atoms in the oxide lattice after the BCP may cause a slightly larger $\Delta$BE of Nb\textsuperscript{+5} component with respect to Nb\textsuperscript{0} owing to high electronegativity of fluorine.
However, currently we did not detect any correlation of $\Delta$BE of Nb\textsuperscript{+5} component with the amount  of fluorine detected. 
Interestingly, a shift of the Nb\textsuperscript{+5} to higher BE was observed upon phosphate-ion adsorption onto niobium in \autocite{pavan2005adsorption}.
It has been shown by us, that the pentoxide thickness of the baked air exposed niobium correlated with the amount of chemically bound carbon, i.e. the more carbide is present, the thinner is the formed Nb$_2$O$_5$ oxide.
Thus, carbon in the lattice also influences the oxide growth, and therefore the Nb\textsuperscript{+5} binding energy.

The niobium RF cavities usually undergo HPR before the thermal processing.
This treatment removes the contamination (phosphate ions, (H$_2$PO$_4$)\textsuperscript{–}, organics, surfactant, etc.) which is adsorbed to niobium surface after the chemical etching.
As it has been mentioned above, as fluorine ions are located within the niobium lattice they can not be removed by HPR.
The HPR treatment increases the thickness of the native oxide, and the XPS data may be misinterpreted and it may be concluded that the fluorine quantity decreased or eliminated by this treatment.
In \autocite{tyagi2012study} it was demonstrated with XPS and SIMS that indeed the concentration of fluorine depleted in the surface layer of \SI{0.5}{\nano\meter}.
Fluorine is known to be outgassed at temperatures above \SI{600}{\degreeCelsius} \autocite{jouve1986x}, so the \SI{800}{\degreeCelsius} anneal applied to cavities for niobium re-crystallization and hydrogen outgassing also mitigates the fluorine-contamination.
Noteworthy, in the standard “mild” (\SI{120}{\degreeCelsius}/48 h) baking, the mid-T baking and also the furnace baking the niobium surface is not \SI{800}{\degreeCelsius}-annealed directly prior to baking and subsequent RF testing.
This means that fluorine is present in the niobium lattice after such treatments and thus may have an impact on the performance of the SRF cavities.
It is especially critical for the mild baking since the amount of fluorine is still pretty large. The mechanism of the improved cavity performance upon the mild baking (mitigation of the so-called high-field $Q$-slope) has been the topic of many investigations during the last three decades \autocite{visentin1999cavity, visentin2010involvement, ciovati2006improved}, and to our best knowledge the fluorine contamination as the lattice impurity has never been discussed.
However, in some works it has been stated that the observed changes in the performance of cavities caused by mild baking could not be explained merely by the change of the diffusion profile of oxygen in niobium at \SI{120}{\degreeCelsius} \autocite{kneisel2000preliminary, antoine2012materials}.

Despite of this, the effects related to physical properties of niobium oxyfluoride compounds (NbO$_x$F$_y$) may also be present.
For example, it has been reported that ReO$_3$-type structures such as NbO$_2$F exhibit negative thermal expansion \autocite{wilkinson2014history}. 
According to our data, in the new baking schemes at \SIrange{200}{300}{\degreeCelsius} the content of fluorine in the state of niobium fluoride or oxyfluoride is even larger than of niobium carbide.
Thus, the influence of fluorine in niobium on the cavity characteristics has to be explored in more detail.
For such purpose niobium has to be treated in a similar way but eliminating the fluorine contamination prior to the thermal treatment by annealing at \SI{600}{\degreeCelsius} or higher.
Alternatively, an HF-free chemical treatment has to be used.
The first option is only possible for the samples meant for measurements in a quadrupole resonator as the cavities are tricky to be prepared mainly due to worse vacuum condition achieved thus making a post-chemical treatment necessary.
The fluorine influence on the superconducting gap via scanning tunneling spectroscopy, on critical fields via SQUID magnetometry, etc. would be interesting to explore. 

\subsubsection{Possible Role of the Remaining Surface Oxides}
\hfill\\ 
After the baking, a mix of niobium oxides with various oxidation states forming the defective distorted niobium lattice is left after the pentoxide dissolution. 
These oxides may have an impact on the cavity performance if the cavity interior was not subjected to air or HPR before the cavity testing.
The oxides possess varying electronic properties.
For example, NbO$_2$ (Nb\textsuperscript{+4}) is electrically characterized as a semiconductor (at room temperature) \autocite{nico2016niobium}. 
A monolayer of Nb$_2$O$_3$ is theoretically predicted to possess quantum anomalous Hall effect \autocite{zhang2017intrinsic} and its properties have been recently explored \autocite{wang2019atomic}.
However, the existence of Nb$_2$O$_3$ in a bulk form is questionable \autocite{brauer1941oxyde}.
For example, it was not observed upon thermal reduction of the anodically oxidized films \autocite{king1990kinetic} and of the films annealed at \SI{2100}{\degreeCelsius} \autocite{delheusy2008x}.
In our experiments the Nb\textsuperscript{+3} state was absent during \SI{200}{\degreeCelsius}/\SI{11.5}{\hour} baking, present at \SI{230}{\degreeCelsius}/\SI{15}{\hour}, and appeared and dissolved at \SI{300}{\degreeCelsius}/\SI{3}{\hour} (only trace amounts of Nb\textsuperscript{+3}  were left) and \SI{400}{\degreeCelsius}/\SI{3}{\hour}.
It is not excluded that Nb-F bonding may also be associated with Nb\textsuperscript{+3} which is normally interpreted as Nb$_2$O$_3$ in the XPS studies.

Further, NbO in a bulk form is a cubic NaCl-like crystal and has the largest amount of ordered vacancies (25\%) in both metal and oxide sublattices among all transition metal oxides \autocite{kurmaev2002electronic}.
This compound is known to possess metallic type of electrical conductivity (with a resistivity of \SI{1.8}{\micro\ohm\meter} at \SI{4.2}{\kelvin} \autocite{hulm1972superconductivity}).
The reported critical temperature of NbO varies between \SI{1.2}{\kelvin} \autocite{khan1974magnetic} and \SI{1.61}{\kelvin} \autocite{okaz1975specific}.
In our samples the percentage of the Nb 3$d$ area referred to NbO (Nb\textsuperscript{+2}) varies between 10-19\% within the probed by XPS information depth for the \SIrange{200}{300}{\degreeCelsius} baked niobium.
This would approximately correspond to a single NbO layer.
Thus, NbO may influence the residual losses of the cavities.

\subsubsection{Kinetics of Oxide Reduction}
\hfill\\ 
The kinetics of native oxide reduction were estimated at the respective baking temperature (without accounting for the sample heating-up stage), providing that the pentoxide was still present at the surface.
Our data fitted well both the first- and second-order reactions determined as $ln(\frac{C}{C_0})$ and $\frac{C_0-C}{C}$ vs. time, respectively.

However, the data did not fit the model $ln(\frac{C-C_e}{1-C_e})$ with the oxygen concentration at the end of the anneal, $C_e$=0.45, used in \autocite{king1990kinetic} in which the dissolution of the anodic niobium oxide film doped with Zr (\SI{1}{\atpercent}) was studied (it was stated that Zr does not affect the kinetics).
The obtained by us values of the activation energy are much lower than in \autocite{lechner2021rf, king1990kinetic} but similar to \autocite{veit2019oxygen}.
In \autocite{veit2019oxygen} the samples represented niobium single-crystals degassed at \SI{2100}{\kelvin}, cleaned with Ar\textsuperscript{+} and subsequently oxidized with O$_2$ gas in a vacuum chamber at room temperature.
However, it is not clear which kinetic model was used for the data fitting in this work.
Based on this, we come to a conclusion that the discrepancy between the data obtained in our work and the data of \autocite{lechner2021rf, king1990kinetic} is not related to the pre-dissolution of Nb$_2$O$_5$ during the heating stage (which was not accounted) or fluorine contamination.
Therefore, the kinetics of oxide reduction has to be further explored in more detail considering the material history, type of oxide and possible contamination. 

The diffusion equation for the continuous plane source \autocite{jaeger1959conduction} was used for the calculation of the oxygen-depth profiles upon oxide-layer dissolution during the bakings at \SIrange{200}{300}{\degreeCelsius}.
The resulted curves demonstrated the distribution of oxygen atoms under the remaining surface oxides and an oxygen-segregated surface layer.
A more thorough analysis would be required exploiting the thermodynamic relations for the estimation of the oxygen concentration profile at \SI{400}{\degreeCelsius}.

\subsubsection{Contamination-free Baking}
\begin{figure} 
	\includegraphics[width=\linewidth]{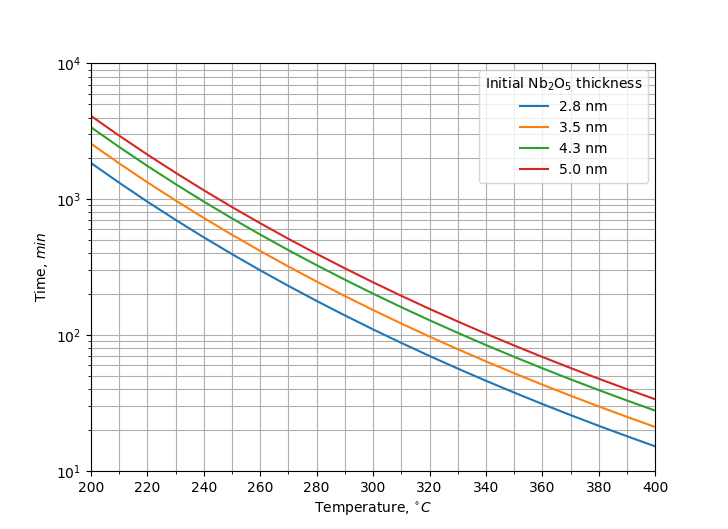}
	\caption{The curves correspond to 1 nm-thick Nb$_2$O$_5$-layer that remains after the baking at the specified temperatures and durations and protects the niobium surface from the interaction with foreign impurities. The curves are given for various initial thicknesses of Nb$_2$O$_5$: \SI{2.8}{\nano\meter} (blue); \SI{3.5}{\nano\meter} (orange); \SI{4.3}{\nano\meter} (green); \SI{5}{\nano\meter} (red).}
	\label{fig:1nm_boundary}
\end{figure}

\hfill\\ 
Since, according to our data, the minimal thickness of the top pentoxide layer protecting the niobium surface from the interaction with impurities (carbon, phosphorus, etc.) is about \SI{1}{\nano\meter}, it is reasonable to calculate the approximate annealing time of the cavity at a specified temperature maintaining this Nb$_2$O$_5$-thickness threshold.
In Figure \ref{fig:1nm_boundary} such curves are shown for the niobium having various initial thickness of Nb$_2$O$_5$: from \SI{2.8}{\nano\meter} (for the BCPed Nb after 2-3 days) to \SI{5}{\nano\meter} (for the cavities after an intensive HPR).
Thus, the cavity performance would not deteriorate for the reason of contamination of niobium surface with the impurities (i.e. the carbides, phosphides, etc.) if the temperature and the duration of baking are chosen below the respective curve.
On the other, for the treatment parameters laying above this curve, phases of Nb-C and Nb-P will form even in UHV environment.
Their amount will depend on temperature, duration and cleanliness of both the niobium sample surface and the furnace.

In  \autocite{tamashevich2023moderate} a vacuum baking of a single-cell niobium cavity at \SI{230}{\degreeCelsius} during 13 h has been demonstrated by us. The treatment resulted in a significant increase of Q$_0$ of the cavity as well as an anti-Q-slope phenomenon and no degradation of the maximal field. At that, the residual resistance decreased from \SI{20}{\nano\ohm} to \SI{9}{\nano\ohm}.

\section{Conclusion}
\label{conclusion}
Employing the synchrotron XPS technique we have thoroughly studied niobium samples subjected to “mid-T baking” and “furnace baking” thermal treatments that have demonstrated high $Q_0$ values on SRF cavities.
It has been established that the chemical composition of the niobium surface treated with BCP is represented by Nb, O, F elements arranged in Nb$_2$O$_{5-x}$/NbO/(Nb + interstitials) structure of the native oxide with a presumable doping with fluorine at least over the depth probed by XPS.
The possible role of fluorine in the educed Nb\textsuperscript{+5} to Nb\textsuperscript{+4} reduction under the impact of an X-ray beam at room temperature and during the thermal treatments has been discussed.
By annealing the niobium samples at \SIrange{200}{400}{\degreeCelsius} in UHV, the kinetics of the native oxide reduction was studied \textit{in-situ}.
The Nb-C and Nb-P chemical states originating from the interaction with surface impurities such as carbonaceous layer and phosphate ions were quantitatively estimated.
Because of their low percentage the precipitates of new phases were neither determined by SEM/EDX nor XRD.

The fluorine that had been loaded during BCP was found to increase its surface concentration during baking and to deplete (evidently by outgassing of Nb-F species) at higher temperatures.
Thus, this finding would help better understand the performance of the cavities treated with "the mild baking" (\SI{120}{\degreeCelsius}/\SI{48}{\hour}) and “the nitrogen doping” as distinguished from the pre-annealed at \SI{800}{\degreeCelsius} (for example, “infusion” with or without nitrogen).

The critical thickness of the Nb$_2$O$_5$ layer of $\approx$\SI{1}{\nano\meter} below which the formation of Nb-C and Nb-P bonds proceeds at a high rate has been determined.
It has been proposed to increase the thickness of the native oxide a few nanometers prior to baking in order to explore the impurity-free mid-T baking on cavities. 
The range of temperature and duration parameters of thermal treatments at which the niobium surface would not be contaminated with impurities is determined, and is of practical importance for the cavity production. 

It has been established that the controlled diffusion of oxygen in niobium when the native-oxide layer represents an oxygen source is realized at temperatures \SIrange{200}{300}{\degreeCelsius}, while at \SI{400}{\degreeCelsius} more oxygen is provided by a residual pressure in the UHV chamber, and thus uncontrolled doping with oxygen (and other impurities) takes place.
This finding is important for interpreting the cavity performance as well as adequate modelling of the diffusion process of solute gas atoms (O, N, C, etc.) in niobium using the solution of the diffusion equation for the continuous plane source. The calculated oxygen depth profiles would be useful for the analysis of SRF properties of similarly treated niobium cavities. Further in-depth studies of the the kinetics of the niobium oxide reduction by oxygen dissolution into niobium are necessary, since the origin of the discrepancies in the obtained activation energies as well as the types of reaction models of the previously reported data are currently not well understood.

The presented information on the baked niobium surface layer after the native oxide dissolution 
would be helpful for interpretation of the results of the cold RF tests of cavities as well as for optimization of their performance.



\section{Acknowledgements}
\label{acknowledgements}

We acknowledge the opportunity to perform measurements at HZB-facilities of the Institute for Silicon Photovoltaics (F. Ruske) and the X-Ray CoreLab (A. Ramírez Caro).
Particularly, I. Rudolph is acknowledged for the access to chemical facilities at the WI-APG department of HZB.
We thank HZB for the allocation of synchrotron radiation beamtime. 
A.M. and D.S. acknowledge the BMBF (grant no. 05K19KER and 0519ODR, respectively).
A.P. has been supported by InnovEEA Project (01.01.2020-31.12.2024).

\clearpage


\printbibliography 


\end{document}